\documentclass[final, 5p, times, twocolumn, 11pt]{elsarticle}
\pdfoutput=1 
\usepackage{url}
\urldef{\ultraxurl}\url{http://www.ultrax-speech.org}

\usepackage{graphicx}
\graphicspath{{figures/}}

\usepackage{csquotes}
\usepackage{xcolor}
\usepackage{textcomp}
\usepackage{gensymb}
\usepackage{amsmath}
\usepackage{amssymb}
\usepackage{booktabs}
\usepackage[ruled,vlined]{algorithm2e}
\usepackage{hyperref}
\hypersetup{
    colorlinks = true,
    linkbordercolor = {white},
    linkcolor = {blue}
}

\usepackage{balance}
\usepackage{lastpage}
\usepackage{tipa}

\journal{Speech Communication}

\biboptions{authoryear}

\begin{document}

\begin{frontmatter}

\title{Automatic audiovisual synchronisation for ultrasound tongue imaging}

\author[a]{Aciel Eshky}\corref{cor1}
\ead{a.eshky@ed.ac.uk}

\author[b]{Joanne Cleland}
\ead{joanne.cleland@strath.ac.uk}

\author[a]{Manuel Sam Ribeiro}
\ead{Sam.Ribeiro@ed.ac.uk}

\author[b]{Eleanor Sugden} 
\ead{eleanor.sugden@strath.ac.uk}

\author[a]{Korin Richmond}
\ead{korin.richmond@ed.ac.uk}

\author[a]{Steve Renals}
\ead{s.renals@ed.ac.uk}

\address[a]{The Centre for Speech Technology Research, University of Edinburgh, UK}
\address[b]{Psychological Sciences and Health, University of Strathclyde, UK}

\cortext[cor1]{Corresponding author}

\begin{abstract}
Ultrasound tongue imaging is used to visualise the intra-oral articulators during speech production. It is utilised in a range of applications, including speech and language therapy and phonetics research.
Ultrasound and speech audio are recorded simultaneously, and in order to correctly use this data, the two modalities should be correctly synchronised. Synchronisation is achieved using specialised hardware at recording time, but this approach can fail in practice resulting in data of limited usability. 
In this paper, we address the problem of automatically synchronising ultrasound and audio after data collection.
We first investigate the tolerance of expert ultrasound users to synchronisation errors in order to find the thresholds for error detection. We use these thresholds to define accuracy scoring boundaries for evaluating our system. %, with the aim of using the thresholds to evaluate our system. 
We then describe our approach for automatic synchronisation, which is driven by a self-supervised neural network, exploiting the correlation between the two signals to synchronise them. We train our model on data from multiple domains with different speaker characteristics, different equipment, and different recording environments, and achieve an accuracy $>$92.4\% on held-out in-domain data.
Finally, we introduce a novel resource, the Cleft dataset, which we gathered with a new clinical subgroup and for which hardware synchronisation proved unreliable. We apply our model to this out-of-domain data, and evaluate its performance subjectively with expert users. Results show that users prefer our model's output over the original hardware output 79.3\% of the time.
Our results demonstrate the strength of our approach and its ability to generalise to data from new domains. 

\end{abstract}

\begin{keyword}
Automatic audiovisual synchronisation \sep synchronisation error tolerance \sep ultrasound tongue imaging 
\end{keyword}

\end{frontmatter}

%\linenumbers
\section{Introduction}

Ultrasound tongue imaging visualises the shape, position, and movement of the tongue during speech production. It is utilised in a number of applications including speech and language therapy, phonetics research, second language learning, and silent speech interfaces \citep{cleland2019enabling, lawson2015seeing, wilson2006ultrasound, hueber2010development}. In the majority of applications, ultrasound is acquired simultaneously with audio, and for the data to be correctly processed and analysed, the two modalities should be correctly synchronised. Synchronisation can be achieved at recording time using specialised hardware \citep{hueber2008acquisition}, however, this approach can fail in practice resulting in data of limited usability \citep{cleland2018personal}. Furthermore, synchronisation information is not always available for historical data \citep{bakst2019post}. While manual synchronisation is possible, it is time consuming, and particularly challenging in the absence of useful audiovisual cues such as stop closures and bursts. %\textcolor{purple}{[ref?]}. 
% manual synchronisation is a lengthy and tedious process \citep{bakst2019post} it's unclear whether manual synchronisation is possible 
%
To address the lack of a mitigation strategy for the failure of hardware synchronisation, we previously introduced a method to automatically synchronise ultrasound and audio after data collection \citep{eshky2019synchronising}, and to our knowledge, no work prior to ours attempted this. Our approach used a self-supervised neural network which exploits correlations between the two signals to synchronise them without the need for manual annotation.  

In this paper, we expand on our previous work. Our first novel contribution is a detailed investigation of the tolerance for synchronisation error by expert ultrasound users. While the tolerance is known for lip video \citep{bt1359relative}, no prior work examines it for ultrasound tongue imaging.
This investigation allows us to identify the threshold for detecting synchronisation error, and to define accuracy scoring boundaries for evaluating synchronisation systems. %which we later use to evaluate our synchronisation system.

Our second contribution builds directly on our previous work in \cite{eshky2019synchronising}. We adopt the UltraSync architecture, retraining the model on data from multiple domains with different speaker characteristics, different equipment, and different recording environments to give it the best chance of generalising to data from new domains, and evaluate the model in the first instance on held-out data of the same domain. %We evaluate in the first instance on held-out data from the same domains.%, achieving good performance on child data ($>$83.6\% accuracy) and adult data ($>$96.1\% accuracy). 

Our final contribution is a novel dataset which we recorded from children diagnosed with cleft lip and palate; a clinical subgroup not previously examined in the context of automatic audiovisual synchronisation, or indeed, automatic processing. Hardware synchronisation proved unreliable for the Cleft data, making it a prime application candidate for our model. Because this data was collected with a new clinical subgroup, in a different environment, and using varied ultrasound settings, we are able to use it to test our model's ability to generalise. We apply our model to this out-of-domain data, and evaluate its performance subjectively with expert users. As part of this work, we make the dataset available to the research community in open format. 
%The results show that users prefer our model to the original hardware output 79.3\% of the time, %and that this result is significant, thus demonstrating the strength of our approach and its ability to generalise to new domains. 

The paper is organised as follows. In Section~\ref{sec:background}, we cover related background on ultrasound tongue imaging and audiovisual synchronisation. 
In Section~\ref{sec:data}, we describe the ultrasound tongue imaging resources we use for our experiments, and introduce our novel dataset, the Cleft data, which was poorly synchronised at recording time. 
In Section~\ref{sec:human_experiment}, we describe the perceptual experiment we designed to identify the threshold for detecting synchronisation errors for ultrasound and audio. We use these thresholds to evaluate our system in the section that follows.
In Section~\ref{sec:model}, we describe our automatic synchronisation system, 
%and the process of training the model driving it. 
then present automatic evaluation on held-out in-domain data. 
In Section~\ref{sec:cleft_sync}, we apply our approach to the Cleft data, and evaluate the output subjectively in a second perceptual experiment. 
We summarise our findings in Section~\ref{sec:conclusion} and conclude with a discussion in Section~\ref{sec:discussion}.

\section{Background}
\label{sec:background}
To put our work in context, we first present background on ultrasound tongue imaging and its main applications.
Then, we transition to audiovisual synchronisation, explaining how it is typically achieved and why it can fail in practice. 
We discuss user tolerance to synchronisation errors,
and present previous research on audiovisual synchronisation,
including work on lip video and how it relates to ultrasound. 

\begin{figure}[t]
\includegraphics[width=\columnwidth]{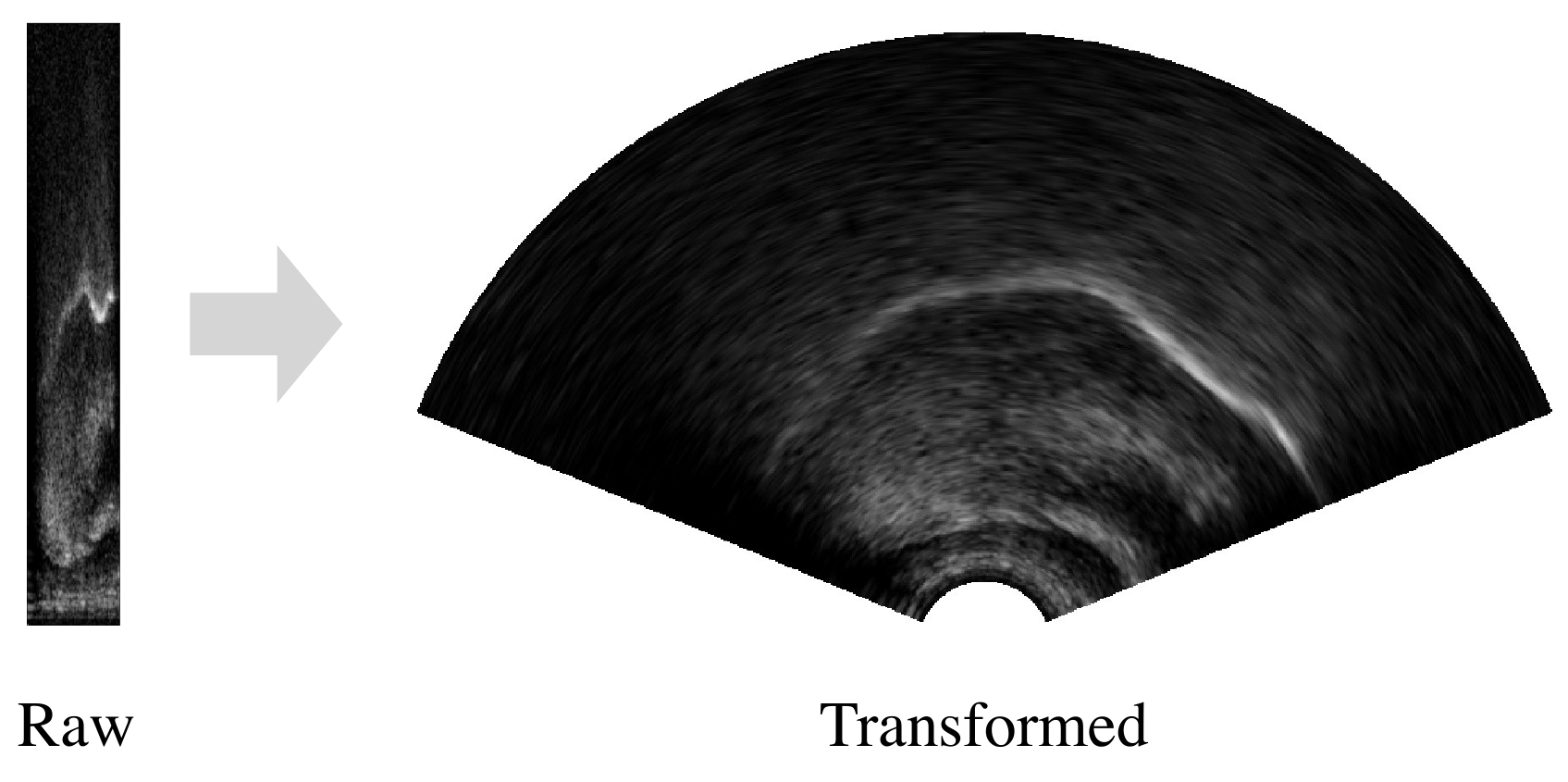}
\caption{Each ultrasound frame is captured as a matrix of raw reflection data (scan lines $\times$ echo returns) and then transformed into real world proportions for visualisation.}
\label{fig:raw_trans}
\end{figure}

\begin{figure}[t]
\includegraphics[width=\columnwidth]{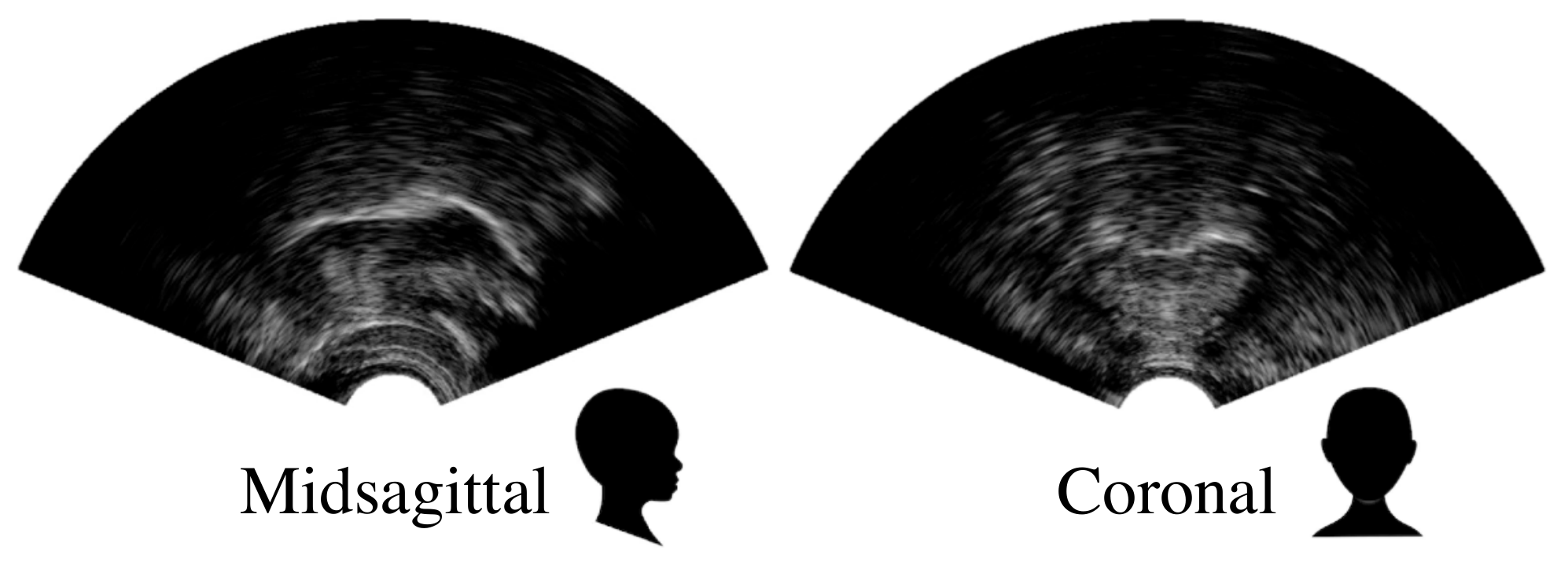}
\caption{Examples of mid-sagittal and coronal ultrasound tongue images for a child (female, aged 5) with bilateral cleft lip and palate (BCLP), taken from the Cleft dataset (speaker 14). The tip of the tongue is to the right in the mid-sagittal image.}
\label{fig:samples}
\end{figure}

\subsection{Ultrasound tongue imaging}
Ultrasound tongue imaging uses diagnostic ultrasound to visualise the tongue surface. 
The ultrasound operates in B-mode (brightness mode) in which a linear array of transducers scans a physical surface and returns a matrix of reflection intensities (scan lines $\times$ echo returns) for each scan. 
Ultrasound data can either be stored efficiently as raw reflection data plus the metadata required to transform it into real world proportions for visualisation, or it can be transformed at recording time and stored as videos. Figure~\ref{fig:raw_trans} shows an example of an ultrasound frame in raw and transformed formats.

To image the tongue, the ultrasound probe is placed under the speaker's chin, capturing either a mid-sagittal or a coronal view of the tongue's surface, depending on the orientation of the probe.
Figure~\ref{fig:samples} shows examples of mid-sagittal and coronal ultrasound tongue images. 
Ultrasound is clinically safe, non-invasive, portable, and relatively cheap \citep{gick2002use, stone2005guide}.

In speech and language therapy, ultrasound tongue imaging can be used to diagnose a range of speech difficulties, and to provide visual biofeedback in therapy for different types of speech sound disorders, including those arising from a cleft lip or palate \citep{sugden2019systematic, roxburgh2015articulation, cleland2020impact}.
During intervention, ultrasound can be used as an objective measure of the patient's progress \citep{cleland2020dorsal}, or to complement verbal feedback and contribute to positive reinforcement \citep{roxburgh2015articulation}. 
Ultrasound also assists annotators in identifying covert articulation errors \citep{cleland2017covert} and has been shown to increase inter-annotator agreement when transcribing the speech of children with cleft lip and palate \citep{cleland2020impact}.

Beyond speech therapy, ultrasound is used in phonetics research to compare tongue-shapes for different phones \citep{davidson2006comparing, lee2014whistled, chen2011analysing, lawson2015seeing, ahn2018role}, or to gain insight into speech production through articulatory-to-acoustic or acoustic-to-articulatory mapping \citep{hueber2011statistical, porras2019dnn}. 
Ultrasound is also used for practical tasks such as second language learning and acquisition \citep{wilson2006ultrasound, gick2008ultrasound, Mozaffari2018Guided}, or to drive silent speech interfaces, which can be used to restore spoken communication for users with voice impairments \citep{denby2004speech, hueber2010development, csapo2017dnn, ji2018updating, ribeiro2021tal}.

To complement this broad range of applications, there is a growing interest in automatically processing and analysing ultrasound, for example, by extracting tongue contours \citep{fabre2015tongue, xu2016robust}, 
animating tongue models \citep{fabre2017automatic, chen2018direct}, 
classifying speech articulation errors \citep{ribeiro2019speaker, ribeiro2021automatic},
and most relevant to our work, synchronising it with simultaneously-recorded audio \citep{eshky2019synchronising}.

\subsection{Audiovisual synchronisation}
\label{sec:sync-background}

While ultrasound tongue imaging can be utilised independently, in the majority of applications it is combined with the simultaneously-recorded audio. To be correctly analysed and used, the two modalities should be correctly synchronised. At recording time, specialised hardware captures the relative time difference between the two signals as an offset in milliseconds, and stores it as metadata \citep{wrench2018sono, wrench2018articulate}. Audio leads if the offset is positive, and lags if negative. Applying the offset to an utterance simply involves cropping the leading signal and the end of the trailing signal.

In practice, hardware synchronisation can fail, either as a result of user error, such as incorrectly connecting and operating devices \citep{cleland2018personal}, or as a result of faulty or inferior hardware components, such as low-quality sound cards \citep{wrench2018personal},
A failure in synchronisation limits the usability of the data \citep{bakst2019post}, and while manual synchronisation is possible, it is time-consuming and challenging, especially in the absence of useful audiovisual cues, such as stops and bursts.

User tolerance for synchronisation error depends on the application. Speech and language therapists mainly use recorded ultrasound for playback in intervention sessions to qualitatively evaluate a patient's performance \citep{cleland2020impact}, and therefore the synchronisation need only be \emph{perceived} as acceptable by the viewer. In contrast, phoneticians often use acoustic landmarks, such as plosive bursts, to annotate articulatory data, in which case synchronisation should be more precise. Because we work mainly with speech and language therapists, we focus in this paper on the former case. 

The majority of research on audiovisual synchronisation focuses on lip videos due to their relevance to broadcasting where synchronisation errors can become objectionable to viewers. In contrast, synchronising audio and ultrasound has received less attention despite its importance. However, because the movement of the articulators (tongue and lips) are correlated \citep{yehia1998quantitative}, we regard prior work on lip synchronisation as relevant. 
% to this work.
A previous study relying on subjective evaluation found that 
lip synchronisation errors between -185ms to 90ms are acceptable to viewers, and that the threshold for error detection is -125ms to 45ms \citep{bt1359relative}. The study also reported that errors in the range of -95ms to 22.5ms are undetectable to viewers. No such study has been conducted for ultrasound, and therefore the thresholds for detecting synchronisation errors are unknown. In this paper, we address this research gap by examining whether lip thresholds also hold for ultrasound. This investigation allows us to refine our evaluation of automatic synchronisation systems.

Some prior work has been dedicated to automating lip synchronisation. 
Older approaches investigated the effects of using different representations and feature extraction techniques on finding dimensions of high correlation \citep{sargin2007audiovisual, bredin2007audiovisual, garau2010audio}. However, these approaches required extensive feature engineering. More recently, neural networks, which learn features directly from input, have been utilised for the task \citep{chung2016out} achieving near-perfect accuracy (99\%) on lip synchronisation according to human evaluators. This approach has since been extended to use different methods for creating training samples \citep{korbar2018cooperative, chung2019perfect} and different model training objectives \citep{chung2019perfect}. 

We previously adopted the original approach from \cite{chung2016out}, modifying it for synchronising ultrasound. Our model achieved an accuracy of 82.9\% for child speech therapy data \citep{eshky2019synchronising}, and 97.7\% for adult speech data \citep{ribeiro2021tal}. 
In this paper, we build directly on our previous work, training our model on data from 
multiple domains with different speaker characteristics, different equipment, and different recording environments, and testing our model's ability to generalise to data from a new domain. %new data, gathered with a new clinical subgroup.
\section{Data}
\label{sec:data}

\begin{table*}[t]
\centering
\resizebox{2\columnwidth}{!}{%
\begin{tabular}{@{}ccccccccc@{}}
\toprule
\textbf{Collection} & \textbf{Dataset} & \textbf{Age} & \textbf{Speech disorder} & \textbf{Environment} & \textbf{Speakers} & \textbf{Stabilisation} & \textbf{Ultrasound settings} 
& \textbf{Hardware Sync}\\
\midrule
UltraSuite & UXTD & Child & None & Research lab & 58 & Metal headset & Consistent & Correct\\
 & UXSSD & Child & SSD & Research lab & 8 & Metal headset & Consistent & Correct\\
 & UPX & Child & SSD & Research lab & 20 & Metal headset & Consistent & Correct\\
\midrule
TaL & TaL1 & Adult & None & Hemi-anechoic chamber & 1 & UltraFit headset & Consistent & Correct\\
    & Tal80 & Adult & None & Hemi-anechoic chamber & 81 & UltraFit headset & Consistent & Correct\\
\midrule
Cleft &  & Child & Cleft & Hospital & 29 & AAA headset or handheld & Varied & Poor\\
\bottomrule
\end{tabular}
}
\caption{Data overview.}
\label{table:data-summary}
\end{table*}

This section describes the data we use throughout the paper. We first present existing ultrasound datasets which we use for our experiments, then introduce the novel Cleft dataset which we collected with a new clinical subgroup. Hardware synchronisation proved unreliable for the Cleft data, making it a prime candidate to automatically synchronise. We explain the challenges associated with this data and why we class it as a new domain. 

Table~\ref{table:data-summary} gives an overview of the data presented in this section. All three resources were recorded in Scotland using the Articulate Assistant Advanced (AAA) software \citep{articulate2010articulate}, which stores ultrasound efficiently in raw format, augmented with the metadata necessary to transform it into real world proportions for visualisation. 

\subsection{UltraSuite repository}
The first existing resource is the UltraSuite repository \citep{eshky2018ultrasuite}, which is a collection of three ultrasound and audio datasets gathered from English-speaking children. The data was recorded by research speech and language therapists in a university laboratory.
The first dataset is Ultrax Typically Developing (UXTD), 
collected with 58 typically developing children. 
The second is Ultrax Speech Sound Disorders (UXSSD), 
collected with 8 children with speech sound disorders. 
The third is UltraPhonix (UPX), collected with 20 children with speech sound disorders. 
The data from UXSSD and UPX was recorded over multiple sessions, including baseline, assessment, therapy, post-therapy, and maintenance. 

Ultrasound was recorded using an Ultrasonix SonixRP machine at $\approx$120fps with a 135\degree ~field of view, and the probe was stabilised using a metal headset. 
Ultrasound frames captured a midsagittal view of the tongue 
with 63 scan lines $\times$ 412 echo returns,
and audio was recorded at 22.05 KHz sampling frequency. Audio recordings contained the speech of both the children and therapists. 
Ultrasound and audio were correctly synchronised at recording time using hardware synchronisation, and this was verified by the researchers who collected the data. 

\subsection{TaL corpus}
The second existing resource is the Tongue and Lips corpus (TaL) \citep{ribeiro2021tal}, which is a collection of ultrasound tongue imaging, lip video, and audio data, recorded with 82 native English-speaking adults. We use the ultrasound and audio for our experiments. 

TaL comes in two parts: TaL1 was recorded with a professional voice talent over the course of 6 days, while TaL80 was recorded with 81 speakers with no voice talent experience. Sessions with the voice talent were approximately 120 minutes long, while sessions with the remaining speakers were approximately 80 minutes long. All recordings took place in a hemi-anechoic chamber, resulting in much better audio quality than UltraSuite. 

Ultrasound was recorded with a Micro system at $\approx$80fps with a 92\degree ~field of view, and the probe was stabilised using the UltraFit stabilising headset \citep{spreafico2018ultrafit}.
Ultrasound frames captured a midsagittal view of the tongue
with 64 scan lines $\times$ 842 echo returns,
and audio was recorded at 48 KHz sampling frequency.
Ultrasound and audio were correctly synchronised at recording time using hardware synchronisation, and this was verified by the researchers who collected the data. 

\subsection{Introducing the Cleft dataset}
The Cleft dataset is a collection of ultrasound and audio data, gathered from children with cleft lip and palate. The data was recorded by research speech and language therapists in a hospital environment. For this dataset, hardware synchronisation was incorrectly recorded, and was perceived as inadequate by the speech and language therapists who collected the data. In Section~\ref{sec:cleft_sync} we use our system to synchronise the data automatically.

The dataset was originally collected for clinical phonetics research \citep{cleland2020impact} and stored in proprietary format. We processed it, and through this work make it available to the research community in open format.
The original data was recorded with 39 English-speaking children, however, only 29 of them gave us consent to share their data (18 male, 11 female). We retain the original speaker IDs for consistency with previous research published on this data \citep{cleland2020impact}, but focus in this paper on the 29 speakers whose data we release.

The children were aged 7-11 years at the time of data collection. Each child had either cleft palate only (CP), unilateral cleft lip and palate affecting one side of the lip and palate (UCLP), or bilateral cleft lip and palate (BCLP) affecting both sides. 
Some speakers had syndromes often associated with cleft lip and palate, including Stickler Syndrome, Treacher Collins Syndrome, and Pierre Robin Sequence. One child had an Adenoidectomy and a Tonsillectomy, and another one had scoliosis at the base of their skull. These medical conditions can lead to additional anatomical differences affecting the mandible, which make it challenging to acquire clear ultrasound images. This, combined with the often more severe nature of speech disorders associated with cleft lip and palate make the data more challenging to automatically process than previous datasets, such UltraSuite and TaL.

The data was recorded over a maximum of two sessions: Assessment and Therapy.
Recordings took place in a hospital, and audio recordings contained the speech of both the children and therapists. The majority of utterances were recorded in the midsagittal view, but some were recorded in the coronal view. We annotated the direction of the probe manually and release the annotation with the dataset. See Figure~\ref{fig:samples} for sample ultrasound images taken from the Cleft data. 

Ultrasound was recorded with a Micro system. The frame rate varied between 80-170 fps, and the field of view varied between 90-80\degree. The number of scan lines varied between 44-64, and the echo returns varied between 842-946.
For the majority of speakers, the probe was stabilised with the AAA headset, but for two speakers (speaker 3 and 12), it was hand-held. 
Audio was recorded at 22.05 KHz sampling frequency.

\begin{table}[t]
\centering
\resizebox{0.8\columnwidth}{!}{%
\begin{tabular}{@{}ccccc@{}}
\toprule
\textbf{Utterance type} & \textbf{Type ID} & \textbf{Sagittal} & \textbf{Coronal} & \textbf{Total}\\
\midrule
Words & A & 303 & 0 & 303\\
Non-words & B & 503 & 73 & 576\\
Sentence & C & 344 & 126 & 470\\
Non-Speech & E & 49 & 43 & 92\\
\midrule
All &  & 1199 & 242 & 1441\\
\bottomrule
\end{tabular}
}%
\caption{The number of utterances of each type in the Cleft dataset.}
\label{table:cleft-summary}
\end{table}

We exported the data from AAA's proprietary format into the same format as UltraSuite and TaL. Four files are associated with each utterance. 
The \textbf{prompt} file is a \emph{.txt} file containing the prompt the child was given and the date and time of the recording.
The \textbf{waveform} is \emph{.wav} file sampled at 22.05 KHz. \textbf{Ultrasound} data is stored as a matrix in a \emph{.ult} file and is accompanied by a \emph{.param} text file containing the metadata, such as frame rate, ultrasound frame size, and original hardware synchronisation offset. We complement this data with exported annotation from speech and language therapists in Praat's TextGrid format.

We categorised utterances into four types according to the prompts.
\textbf{Words} contain a group of words designed to sample consonants in different vowel contexts within real words (e.g., ``a core, a sip, a cop, a tool"). 
\textbf{Non-words} are designed to elicit specific phones but are not real English words (e.g., ``acha" for %/t$\int$/)
\textipa{/\textteshlig/}). Many of these utterances contain multiple repetitions of the the same word (e.g., ``acha acha acha acha").
\textbf{Sentences} are designed to examine specific phones in different contexts at the sentence level (e.g., ``Tiny Tim is putting a hat on" for the phone /t/). 
And finally, \textbf{non-speech} utterances are swallowing motions recorded to trace the hard palate. 
We append the type ID to the utterance name (e.g., ``001E.wav").
Table~\ref{table:cleft-summary} summarises the data.

\subsection{Cleft data challenges }
A number of factors make the Cleft dataset more challenging to automatically process than TaL and UltraSuite, leading us to class it as a \emph{new domain}. 
Firstly, the data was recorded with a clinical subgroup with severe speech disorders making audio more challenging to understand than the disordered subset of UltraSuite (UPX and UXSSD).
Cleft patients also exhibit abnormal lingual articulatory patterns which are captured in ultrasound \citep{zharkova2013using}, and which will be different to patterns exhibited in UltraSuite and TaL.
Furthermore, the anatomical differences arising from cleft lip and palate, as well as the additional syndromes that affect some of the children, can give rise to differences in the ultrasound data and in some cases make it more challenging to acquire clear data in the first place. 
 
%acquiring clear ultrasound data more chaThese anatomical differences also made It was also challenging in some cases to acquire clear ultrasound data due to anatomical differences 
%
Secondly, the data was recorded in a hospital environment with a lot of background noise. In contrast, UltraSuite was recorded in a quieter research laboratory, while TaL was recorded in a silent hemi-anechoic chamber. 

Finally, the ultrasound in the Cleft data was recorded at varied settings including different frame rates, fields of view, scan lines, and echo returns, compared to the UltraSuite and TaL datasets which were consistent across speakers. Furthermore, the ultrasound probe was not always stabilised with a headset, leading to further inconsistency in the data. For these reasons we class the Cleft dataset as a new domain. 

Because the Cleft data was poorly synchronised at recording time, we restrict its use to Section~\ref{sec:cleft_sync} where we automatically synchronise it using our system. In the next section, we examine the tolerance of expert users to synchronisation errors. 

\section{Identifying the detection threshold}
\label{sec:human_experiment}

This section aims to identify the threshold at which a synchronisation error becomes detectable to experienced ultrasound users. Identifying this threshold allows us to refine our approach for evaluating our system in Section~\ref{sec:model}.
Because the movement of the articulators (the tongue and lips) are correlated, we turn to a study carried out with human participants which reports 6 different thresholds for lip synchronisation \citep{bt1359relative}. We test whether the lip thresholds also apply to ultrasound in perceptual experiment, which we describe below\footnote{This study was certified according to the Informatics Research Ethics Process (ref no 2019/43362).}. 

\subsection{Experiment}

\begin{table}[t]
\centering
\resizebox{1\columnwidth}{!}{%
\begin{tabular}{@{}cccc@{}}
\toprule
\textbf{Participant ID} & \textbf{Profession} & \textbf{Native English} & \textbf{Dialect}\\
\midrule
1 & SLT & Yes & Scottish \\
2 & Speech Scientist & Yes & British other \\
3 & Speech Scientist & Yes & Scottish \\
4 & Speech Scientist & No & Fluent, non-native \\
5 & SLT & Yes & Scottish \\ 
6 & SLT & Yes & Non-British \\ 
7 & Speech Scientist & No & Fluent, non-native \\
8 & SLT & Yes & British other \\
9 & Speech Scientist & No & Fluent, non-native \\ 
10 & Speech Scientist & No & Fluent,  non-native \\ 
\bottomrule
\end{tabular}
}%
\caption{Details of the participants.}
\label{table:par_1}
\end{table}

The purpose of this experiment was to discover how sensitive experienced ultrasound users are to different synchronisation errors. 
To this end, we recruited a number experienced ultrasound users, 
and asked them to assess the quality of audiovisual synchronisation in a series of recordings. 
During the experiment, we gave each participant pairs of videos containing ultrasound tongue imaging and the corresponding audio, 
and asked them to choose the videos which they perceive to be better synchronised.
Each pair of videos were identical apart from the synchronisation offset. 
For one of the videos, we use the \textbf{correct hardware synchronisation offset}.
For the other video, we \textbf{added an error to the correct offset}.  
The order of the videos was randomised, and the correct choice was unknown to the participants. 
We asked the following question: ``In which of the two videos are the audio and tongue motion better synchronised, A or B?", and gave 3 choices: ``Video A", ``Video B", and ``No perceived difference". We refer to the last as option C. We encouraged participants to make a choice between videos A and B, and to reserve option C for the most challenging cases.
In this setting, the smaller the error the more challenging the task, and therefore, we expect the accuracy of choice to approach 50\% when the error is imperceptible, and 100\% when the error is perceptible. 

The experiment was computer-based, and the videos were displayed on the participants' screens. 
The overall experiment lasted 30-40 minutes, and participants were allowed to complete it over multiple sessions.
All utterances were in the midsagittal orientation with the tip of the tongue to the right.
Ultrasound tongue imaging users typically playback ultrasound at three possible speeds: 1.0$\times$, 0.5$\times$, and 0.25$\times$. We replicated this setting by giving our participants the option to play the videos at these three speeds. 
Participants were required to play each video at least once and up to 6 times at any speed, and could only move to the next video after they had submitted a judgement. 
To qualify, each participant was required to be a fluent English speaker, and either a speech and language therapist or a speech scientist with experience working with ultrasound tongue imaging. We recruited 10 participants whose details we outline in Table~\ref{table:par_1}.

\subsection{Data preparation}

To test synchronisation errors, we required correctly-synchronised data. 
We therefore used the typically developing subset of UltraSuite, UXTD, which was correctly synchronised at recording time using hardware synchronisation. %UXTD contains 58 speakers in total. 
We chose this subset of UltraSuite to avoid distracting our participants with speech sounds disorders. 
The TaL corpus was not used for this experiment, as it was still in the process of being collected. 
%The TaL corpus was not yet available and was therefore not used for this experiment. 

To get a rough idea of the audio quality, we listened to a small sample of audio recordings from each of the 58 speakers, then retained 42 speakers with the fewest interruptions from the therapists, fewest hesitations, and fewest deviations from the prompts. 
We sorted the speakers by the number of utterances, then by the standard deviation of the duration of utterances and chose the top 13 speakers
%a subset that is roughly balanced by gender 
(6 female, 7 male).
%The speakers are: 01M, 05M, 06F, 07F, 09F, 13F, 17M, 19M, 20M, 22M, 23F, 27M, 30F.
These were speakers 1, 5, 6, 7, 9, 13, 17, 19, 20, 22, 23, 27, and 30.
We selected a variety of prompts excluding coughs, and swallows
%, and non-English articulations, 
and limited our selection to utterances shorter than 7.5 seconds. In total, we ended up with 520 unique recordings.

\begin{figure}[t]
\includegraphics[width=\columnwidth]{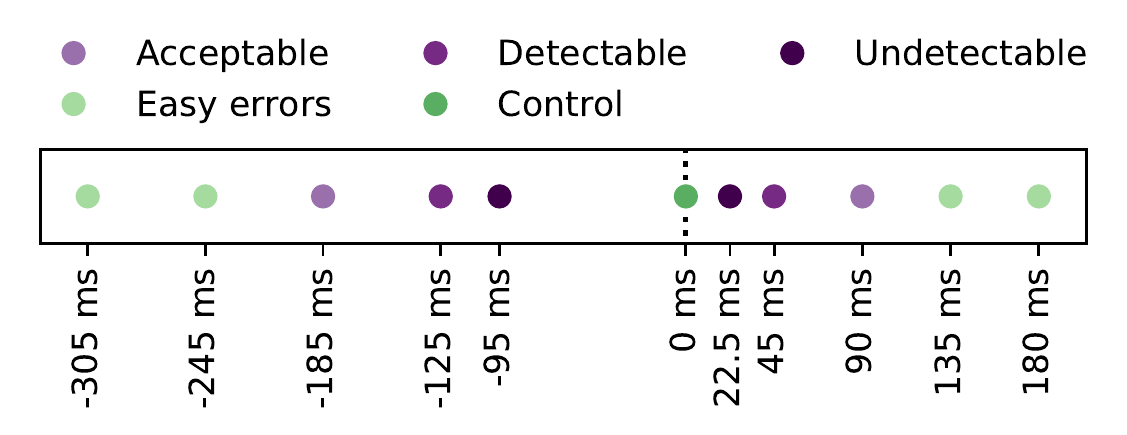}
\caption{\label{fig:errors}
The set of synchronisation errors tested in our experiment. 
Audio lags when the error is negative and leads when positive.
We tested the standard lip synchronisation error thresholds from \cite{bt1359relative} (acceptable, detectable, undetectable). 
The asymmetry indicates that errors are more challenging to detect when audio lags (negative) and easier to detect when audio leads (positive). We further tested four easy errors and add a control of zero error.}
\end{figure}

Next, we selected the set of errors to test, using the thresholds for lip synchronisation from \cite{bt1359relative}. 
\textbf{Lip synchronisation errors} are classed as: 
\begin{enumerate}
    \item \textbf{Acceptable}: between -185ms and 90ms
    \item \textbf{Detectable}: at -125ms and at 45ms
    \item \textbf{Undetectable}: between -95ms and 22.5ms
\end{enumerate}
Note the asymmetry in these thresholds: the magnitude of each positive error is smaller than its negative counterpart, indicating that errors are easier to detect if the audio leads, and more challenging to detect if the audio lags.
 
In earlier iterations of the experiment, we discovered that these thresholds were very challenging for our participants. Therefore, to make the experiment more engaging, and to give the participants less challenging cases to calibrate their answers to, we added four larger errors, two positive, and two negative.
We selected these errors by computing the difference between the detectable and acceptable lip error thresholds, and used this difference to create two new evenly-spaced errors. We did this independently for positive and negative errors. 
Finally, we added a control, an error of zero. In this case, there was no difference between the pair of videos. The reason for adding this case is to test whether there is a bias towards choice A or choice B. For example, always preferring the video at the top of the screen would be a kind of bias.
The final set of errors we tested is: [-305, -245, -185, -125, -95, 0, 22.5, 45, 90, 135, 180] ms (illustrated in Figure~\ref{fig:errors}). 

To create samples for our experiment, we randomly assigned the errors to the utterances.
We drew 500 utterances from our pool of 520 and distributed them among the errors.
We assigned each error 50 unique utterances, with the exception of the two most challenging errors, -95 and 22.5 (undetectable lip error) which we assigned only 25 each to avoid frustrating participants.
Each participant evaluated 60 samples, 40 unique to them, and 20 shared with another participant to allow us to calculate pairwise agreement.
In total, 500 unique samples were evaluated: 400 by a single participant, and 100 twice by a pair of participants, bringing the number of samples to 600.
Each participant evaluated the same number of samples for each error. 
We report the results below.

\subsection{Results}

\begin{table}[t!]
\centering
\resizebox{0.8\columnwidth}{!}{%
\begin{tabular}{@{}cccc@{}}
\toprule
\textbf{Subset} & \textbf{Samples} & \textbf{Accuracy} & \textbf{CI} \\
\midrule
All & 600 & 74.0\% & (70.5, 77.5) \\
Excluding control & 540 & 78.3\% & (74.9, 81.8) \\
A is correct & 262 & 77.9\% & (72.8, 82.9) \\
B is correct & 278 & 78.8\% & (74.0, 83.6) \\
Control (C is correct) & 60 & 35.0\% & (22.9, 47.1) \\
\bottomrule
\end{tabular}
}%
\caption{Overall accuracy of participant responses. CI are 95\% binomial confidence intervals. Control pairs have a zero error for both videos. The percentages of A and B choices are similar indicating no bias towards A or B.}
\label{table:overall-acc}
\end{table}
\begin{table}[t!]
\centering
\resizebox{0.75\columnwidth}{!}{%
\begin{tabular}{@{}cccc@{}}
\toprule
\textbf{Error sign} & \textbf{Samples} & \textbf{Accuracy} & \textbf{CI} \\
\midrule
Negative & 270 & 78.9\% & (74.0, 83.8)\\
Zero (control) & 60 & 35.0\% & (22.9, 47.1) \\
Positive & 270 & 77.8\% & (72.8, 82.7) \\
\bottomrule
\end{tabular}
}%
\caption{Overall accuracy of participant responses by sign. CI are 95\% binomial confidence intervals. The accuracy is symmetrical despite the errors being asymmetrical indicating that the lip synchronisation asymmetry also holds for ultrasound tongue imaging. Negative: audio lags. Positive: audio leads. Zero: no error (control).}
\label{table:acc-by-sign}
\end{table}

The first results are shown in Table~\ref{table:overall-acc}. The overall accuracy of participant choice was 74.0\%. For control questions, where both videos had no synchronisation error, participants selected C only 35.0\% of the time. As for non-control questions, participants chose the correct answer 78.3\% of the time. The percentages of A and B choices were balanced (77.9\% and 78.8\%) indicating no bias in choice towards A or B. Table~\ref{table:acc-by-sign} displays the accuracy by sign. The table shows that accuracy is symmetrical despite the errors being asymmetrical, indicating that the asymmetry that holds for lip synchronisation also holds for ultrasound tongue imaging.

\begin{figure}[t]
\includegraphics[width=\columnwidth]{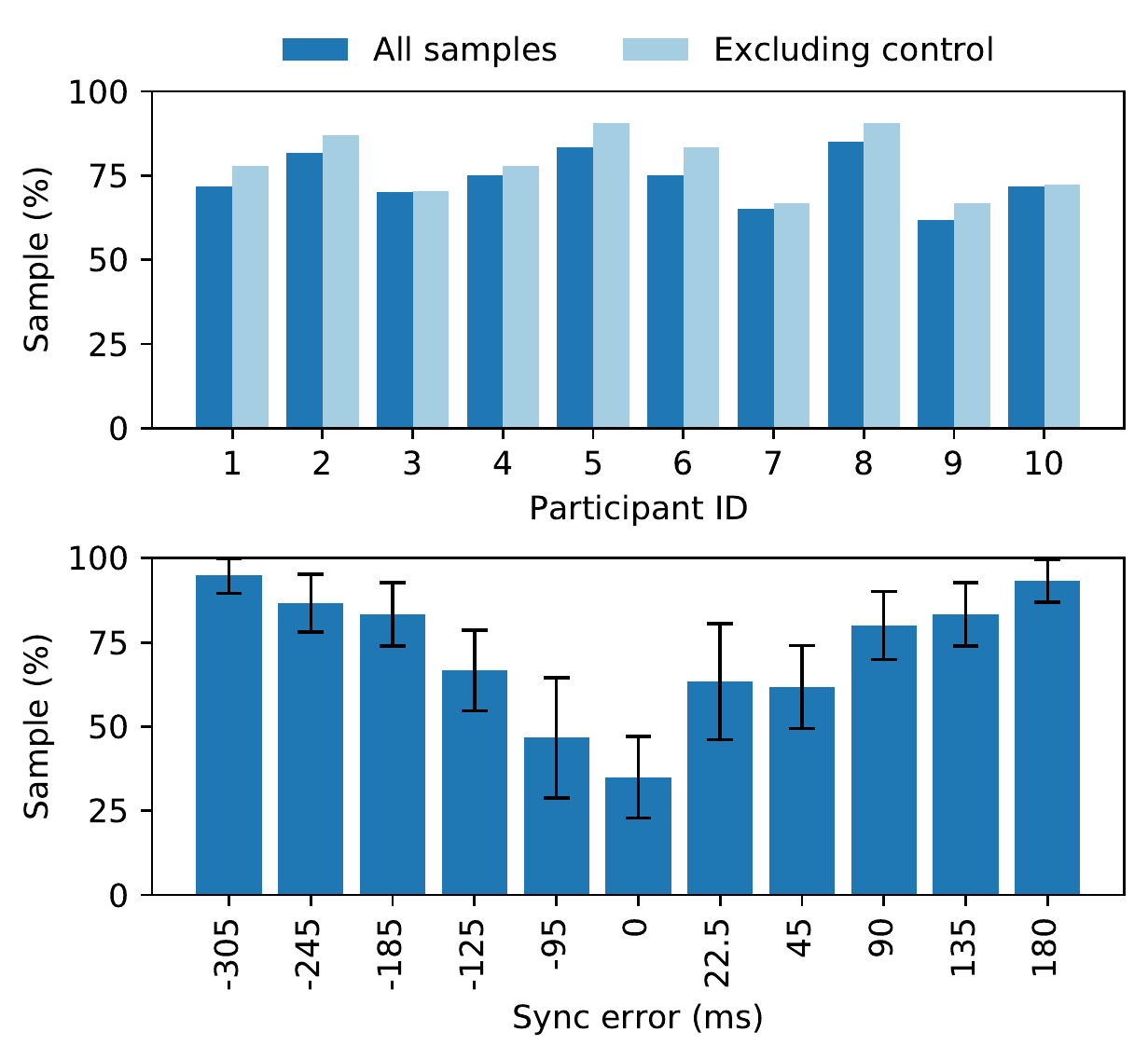}
\caption{\label{fig:accuracy} 
Accuracy of choice shown by participant (top) and by synchronisation error (bottom). 
The smaller the error, the more challenging the task.
The confidence intervals for the undetectable lip errors cross 50\% indicating that they are also undetectable for ultrasound.
The confidence intervals for 45ms also reaches 50\%, indicating that this threshold for detecting lip error is not applicable to ultrasound.
The accuracy at the threshold for acceptable lip error is more reliable.
}
\end{figure}

\begin{figure}[t]
\includegraphics[width=\columnwidth]{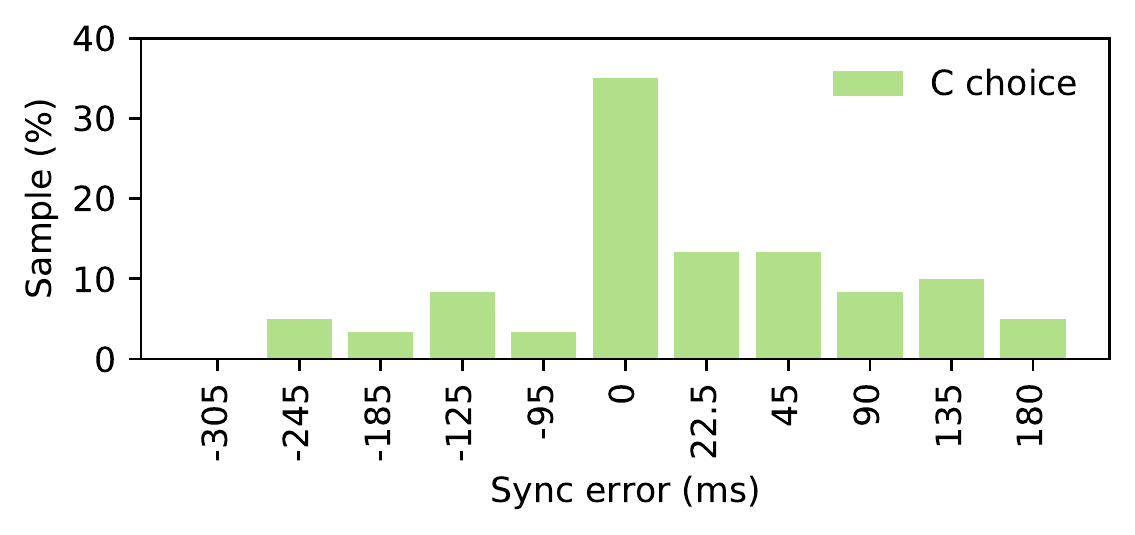}
\caption{\label{fig:c_choice} 
The distribution of C choices (``no perceived difference") per synchronisation error.
The majority of C choices are concentrated at zero error (where there is in fact no difference), 
and the distribution tapers as the errors become larger. 
}
\end{figure}

Figure~\ref{fig:accuracy} breaks the accuracy down by participant and by error. As expected, the smaller the error, the more challenging the task. The confidence intervals for the undetectable lip error thresholds both cross 50\%. The confidence intervals for 45ms reaches 50\%, indicating that even the detectable lip error thresholds are too challenging for ultrasound tongue imaging. We start to see more reliable accuracy at the acceptable lip error thresholds. Finally, Figure~\ref{fig:c_choice} shows the percentage of C choices, or ``no perceived difference". 

\begin{figure}[t]
\includegraphics[width=\columnwidth]{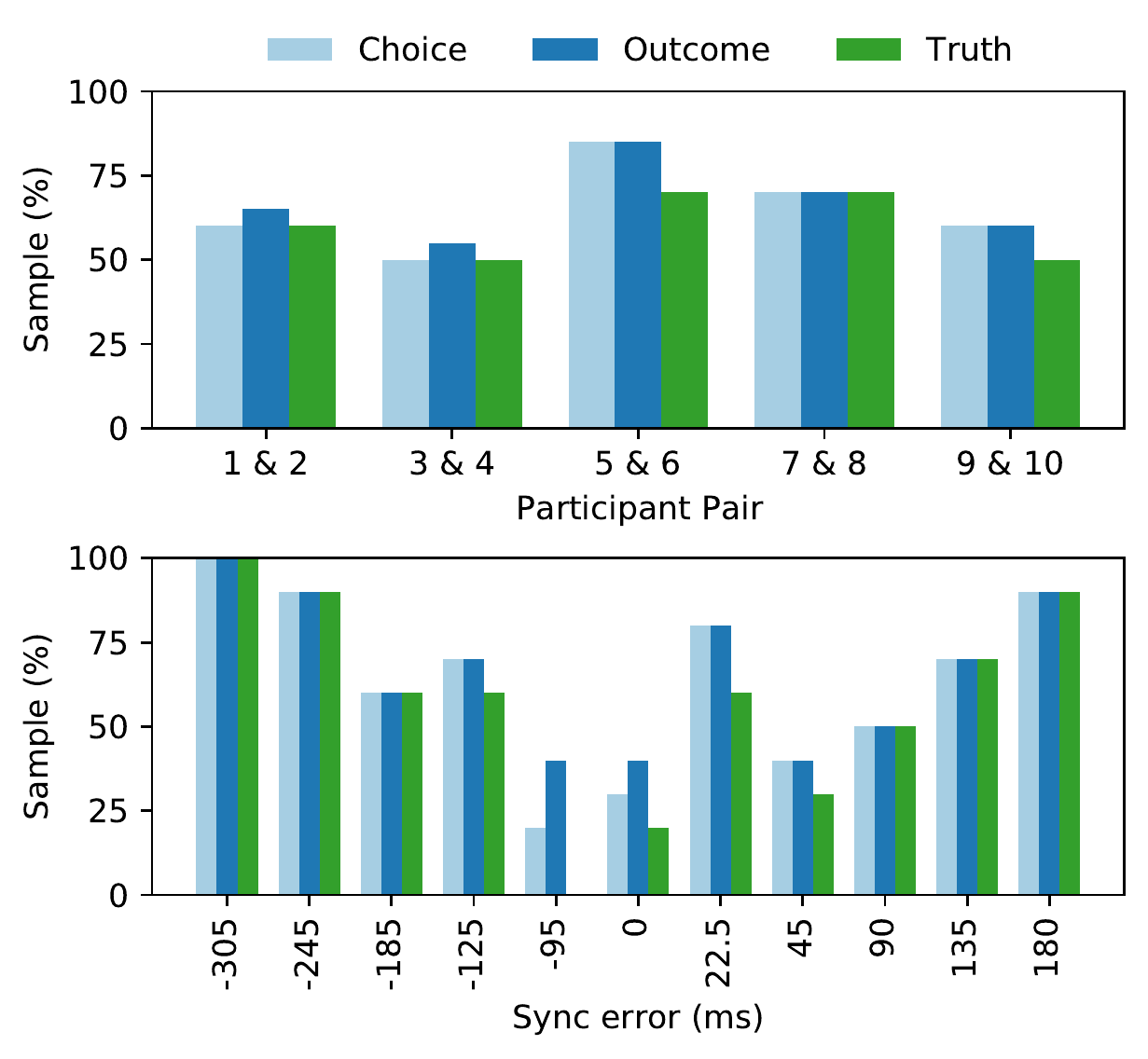}
\caption{\label{fig:agreement} 
Pairwise agreement shown by participant pair (top) and by synchronisation error (bottom). 
Each participant pair shared 20 samples and agreed on at least half of them.
Each error had 20 samples except errors -95ms and 22.5ms which had only 10 each.
The smaller sample size might explain why agreement is lower than
expected for -95ms and higher than expected for 22.5ms.
Otherwise, agreement positively correlates with error magnitude.
}
\end{figure}

Next, we calculated pairwise agreement. Each pair of participants (1 \& 2, 3 \& 4, \ldots etc.) received a common subset of 20 samples. The synchronisation errors had an equal number of common samples, 20 each, with the exception of undetectable lip errors, -95 and 22.5 which had 10 samples each. We calculated the following \textbf{scores for pairwise agreement}: 
\begin{enumerate}
    \item \textbf{Agreement of choice}: did the participants make the same choice (A, B, or C)?
    \item \textbf{Agreement of outcome}: did their choice have the same outcome (both correct or both incorrect)?
    \item \textbf{Agreement with truth}: did the choice match the truth (both correct)?
\end{enumerate}
Figure~\ref{fig:agreement} shows the results by participant pair and by synchronisation error. All pairs of participants agreed on at least 50\% of samples. As expected, the smaller the error, the lower the agreement, with the exception of the undetectable error at -95ms and 22.5ms, where agreement is lower than expected at -95ms and higher than expected at 22.5ms, possibly due to the smaller sample size. 
Another contributor could be the randomisation procedure: because utterances were randomly assigned errors, it is possible that certain errors had easier / more challenging utterances by chance.

%
%

%%%% New results
%
%
\begin{table}[t]
\centering
\resizebox{.75\columnwidth}{!}{%
\begin{tabular}{@{}ccccc@{}}
\toprule
\textbf{Category} &
\textbf{P} & 
\textbf{N} & 
\textbf{Accuracy} &
\textbf{CI} \\
\midrule
SLT                 & 4 & 216 & 85.6\% & (81.0, 90.3) \\
Speech Scientist    & 6 & 324 & 73.5\% & (68.6, 78.3) \\
\midrule
Fluent, non-native  & 4 & 216 & 70.8\% & (64.8, 76.9) \\
Scottish            & 3 & 162 & 79.6\% & (73.4, 85.8)\\
British other       & 2 & 108 & 88.9\% & (83.0, 94.8) \\
Non-British         & 1 & 54  & 83.3\% & (73.4, 93.3) \\
\bottomrule

\end{tabular}
}%
\caption{Accuracy of participant responses excluding control,
broken down by the participants’ professions (top) and their dialects (bottom). 
P is the number of participants, while N is the number of samples. CI are 95\% binomial confidence intervals.}
\label{table:acc_profession_lang}
\end{table}

\begin{table}[t]
\centering
\resizebox{1\columnwidth}{!}{%
\begin{tabular}{@{}cccccc@{}}
\toprule
\textbf{English Speaker} &
\textbf{Profession} & 
\textbf{P} &
\textbf{N} & 
\textbf{Accuracy} &
\textbf{CI} \\
\midrule
Native      & SLT               & 4 & 216 & 85.6\% & (81.0, 90.3) \\
Native      & Speech Scientist  & 2 & 108 & 78.7\% & (71.0, 86.4) \\
Non-native  & Speech Scientist  & 4 & 216 & 70.8\% & (64.8, 76.9) \\

\bottomrule
\end{tabular}
}%
\caption{Accuracy of participant responses excluding control, split by the combination of native language and profession. CI are 95\% binomial confidence intervals. Native English-speaking SLTs perform the task better than non-native English-speaking speech scientists. The CI of the middle group (native English speaking speech scientists) overlaps with the two other groups.}
\label{table:acc_native_prof}
\end{table}
To understand why the overall accuracy varied by participant, we broke the results down by their professions and dialects in Table~\ref{table:acc_profession_lang}.
Four participants were speech and language therapists (SLTs) while six were speech scientists. As for their dialects, 4 were fluent non-native English speakers and 6 were native English speakers: 3 Scottish, 2 non-Scottish British, and 1 non-British. Table~\ref{table:acc_profession_lang} shows that SLTs achieved higher accuracy than speech scientists, 
however, Table~\ref{table:acc_native_prof} shows that the profession of participants co-varied with their native language, and that not all combinations are represented in out data. For example, all non-native English speakers were speech scientists and none were SLTs. While such characteristics may have an effect on a user's sensitivity to synchronisation offsets, from our data it is not possible to isolate the individual effects of profession and native language. 

%however, Table~\ref{table:acc_native_prof} examines the relationship between the profession and native language, showing that 4 out of the 6 speech scientist were non-native English speakers, whereas all SLTs were native English speakers. These results indicate that the native English-speaking SLTs perform the task better than the non-native English-speaking speech scientists, however, it is difficult to draw conclusions from only 10 participants with co-varying characteristics. Further experiments can examine these characteristics in a more controlled fashion.
%%%%

%
%
\begin{figure*}[t!]
\centering
\includegraphics[width=1\textwidth]{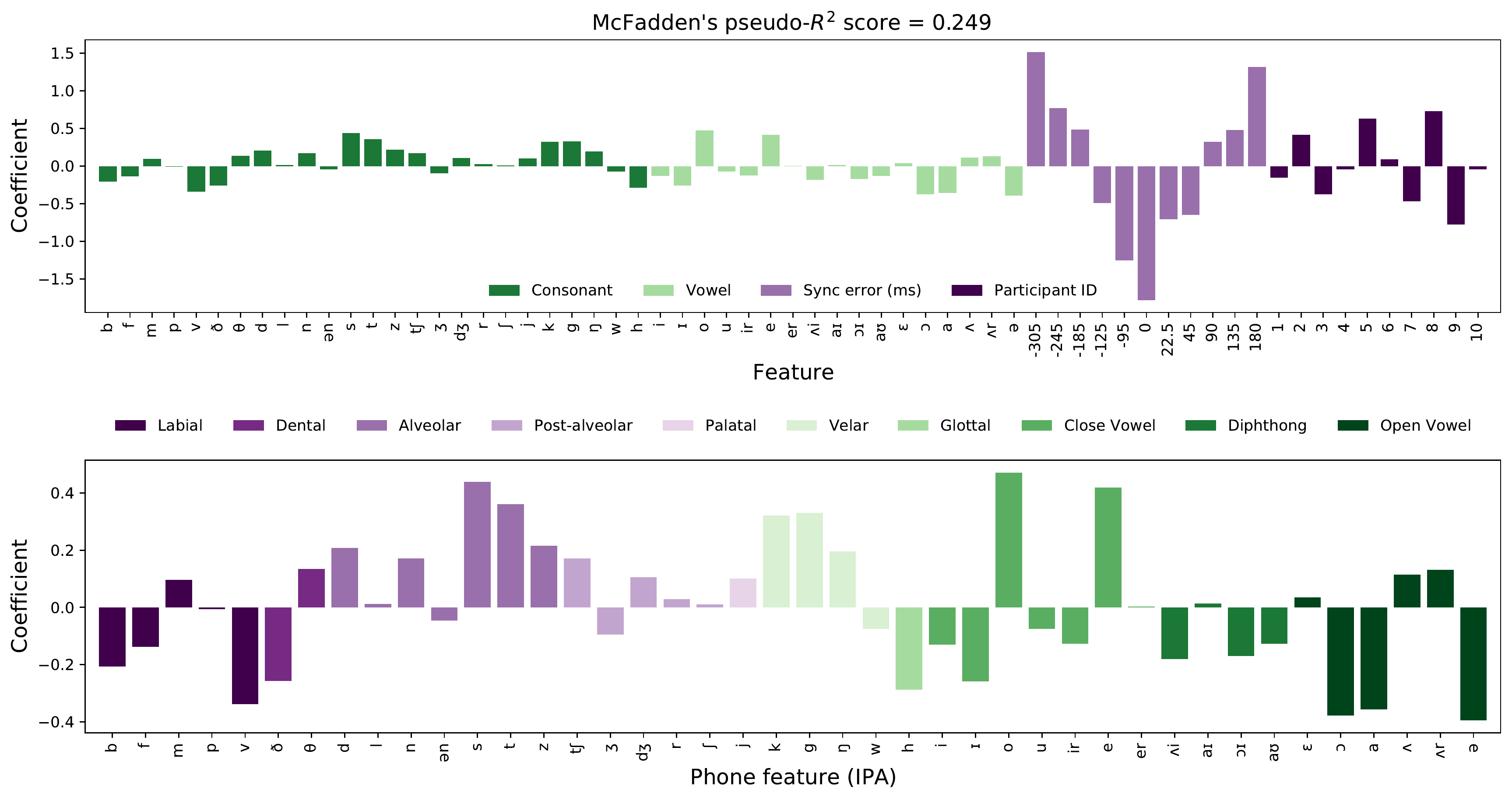}
\caption{\label{fig:linear_model}
The coefficients of a logistic regression model predicting choice outcome (correct / incorrect) from the features shown. 
McFadden's pseudo-$R^2$ score shows the proportion of model variation explained by the features.
\textbf{Top:} the complete set of features. Synchronisation errors falling between $-$125ms and 45ms negatively correlate with a correct choice, while errors $<=-$185ms and errors $>=$90ms positively correlate with a correct choice. Sensitivity to synchronisation errors varies by participant.
\textbf{Bottom:} the phone features colour-coded by the place of articulation. Alveolars, post-alveolars, palatals, and velars positively correlate with a correct choice, while labials, dentals, and glottals negatively correlate with a correct choice. Most vowels (close vowels, open vowels, and diphthongs) negatively correlate with a correct choice, with a few exceptions such as `o' and `e'.}
\end{figure*}

Finally, we conducted a linear analysis, predicting the outcome of choice (correct/incorrect) from the synchronisation error while controlling for the participant and utterance content. 
We represented errors and participants as one-hot encoding vectors, and introduced content features at the phone level to test whether synchronisation errors are easier to detect in the presence of certain phones. 
To map each word to its pronunciation, we used the UXTD pronunciation dictionary supplied with the data. The pronunciation dictionary was compiled for a Scottish accent (to match the accent in the data) using the Combilex lexicon \citep{richmond2009robust, richmond2010generating}.
We found 42 unique phones in the test utterances. 
For each utterance, we created a feature vector of size 42, 
and counted the number of occurrences for each phone. 
For words with multiple pronunciations, we added fractional counts for each phone as $Count=\frac{1}{P}$, where $P$ is the number of pronunciations. 

We then fit a logistic regression model predicting the binary outcome (correct / incorrect) from 63 features: 11 errors, 10 participants, and 42 phones. We used LBFGS with $L2$ regularisation. Upon convergence, the model achieved a log loss of 0.456. To calculate the proportion of model variation that is explained by the features, we used McFadden's pseudo-$R^2$. 
The score falls between 0 and 1, however, in practice, scores ranging from 0.2 to 0.4 are considered excellent \citep{hensher1979behavioural} and indicate that a large proportion of the model is explained by the features. Our model's pseudo-$R^2$ score is 0.249. 

The model coefficients are shown in Figure~\ref{fig:linear_model}. 
The direction of the coefficients (positive / negative) is the direction of correlation with correctness of participant choice.
We find that the tolerance for synchronisation error varies by participant. 
Synchronisation between -125 and 45 ms negatively correlate with a correct choice. For lip synchronisation, these are the thresholds for detection. However, these results indicate that for ultrasound, undetectability extends to this range. Errors $<=$-185 ms and errors $>=$90 ms positively correlate with a correct choice. 
We therefore define the following thresholds for \textbf{ultrasound synchronisation errors}:
\begin{enumerate}
    \item \textbf{Detectable}: at -185ms and at 90ms
    \item \textbf{Undetectable}: between -125ms and 45ms
\end{enumerate}
and use them in Sections~\ref{sec:model} to evaluate our system.

Because we represented phone as fractional counts, and represented participants and errors as one-hot vectors, the magnitudes of coefficients are not directly comparable. However, the direction of the coefficients is the direction of correlation. The results for utterance content meet our expectations. Phones that involve little tongue movement, such as those produced using the lips (for example /b/) or the glottis (for example /h/), negatively correlate with a correct choice. In contrast, phones involving more tongue activity (alveolars, post alveolars, palatals, and velars) positively correlate with a correct answer. This result is intuitive and meets our expectations.

\subsection{Discussion and summary}

In this section, we applied the standard lip error thresholds to ultrasound and tested them in a perceptual experiment with expert ultrasound users. We concluded that detecting synchronisation errors in ultrasound tongue imaging is more challenging than in lip videos. This is perhaps not surprising given that most humans are exposed to audiovisual perception of lip movement from birth, therefore accumulating thousands of hours of experience seeing synchronised lips and audio. The same does not hold for ultrasound; even the most experienced ultrasound users only have tens or hundreds of hours of experience working with synchronised ultrasound and audio. Moreover, ultrasound images, unlike videos of lips, are not a facsimile, or indeed even a video, instead they are a representation of tongue-movements based on echos of high-frequency sound waves and as such are susceptible to artefacts. It is therefore reasonable for the synchronisation error detection threshold to be larger than for lip videos. 

We further concluded that the sensitivity to synchronisation errors varies by participant, after taking into account the linguistic content of utterances and the offsets as co-variates in the linear model. 

Finally, we concluded that sounds involving high tongue activity positively correlate with synchronisation error detection, while sounds involving low tongue activity negatively correlate with synchronisation error detection. 

\section{Automatic synchronisation system}
\label{sec:model}

This section details our approach for automatically synchronising ultrasound and audio. We build directly on our model from \citet{eshky2019synchronising} reiterating its description below. 
We then describe a new experiment, 
introduce two evaluation scores based on the results from Section~\ref{sec:human_experiment}, and present our results on in-domain data. 

\subsection{Model}
We use the UltraSync architecture from \citet{eshky2019synchronising} which previously extended the work of \citet{chung2016out} on lip synchronisation, modifying for ultrasound tongue imaging. The system accepts as input an ultrasound signal and an audio signal, and requires the range of possible offsets to be specified. From this range, the system selects the offset that minimises the distance between the two signals. 

\begin{figure}[t]
\includegraphics[width=\columnwidth]{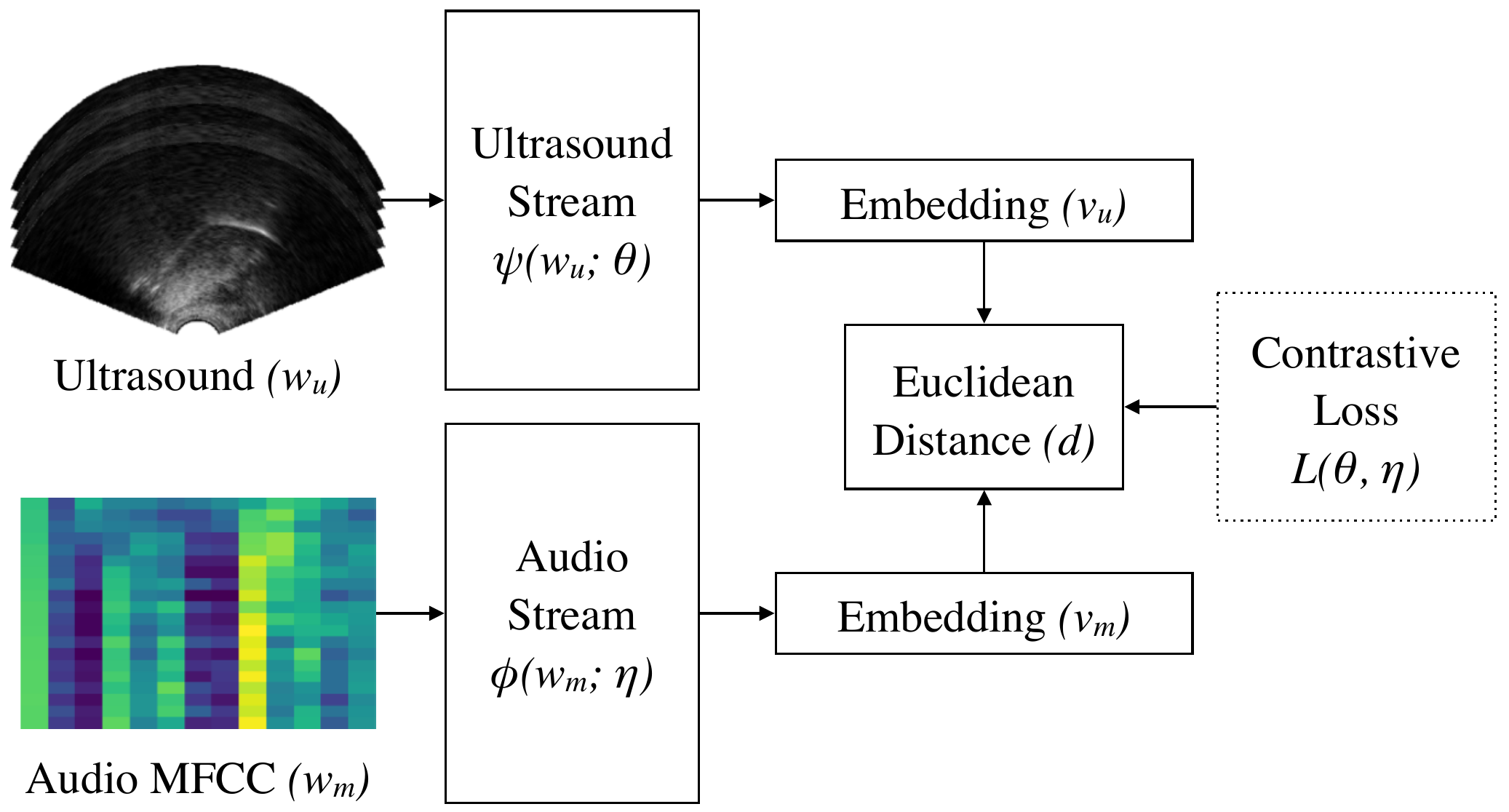}
\caption{The UltraSync model accepts as input a window of ultrasound and a window of audio, represented as MFCC features. Each stream is a convolutional neural network mapping the inputs to low dimensional embeddings. The model then outputs the Euclidean distance between the embeddings, and a contrastive loss function minimises the distance for true pairs and maximises it for false pairs.}
\label{fig:network}
\end{figure}

At the heart of the system is a neural network with two streams, illustrated in Figure~\ref{fig:network}. 
The first stream accepts a short window of ultrasound, and the second accepts a short window of audio. 
The inputs are of different sizes and are high-dimensional. The network maps the pair of inputs to a pair of low-dimensional embeddings of equal lengths, such that the Euclidean distance between them is small when they correlate and large otherwise. 

The learning objective is a contrastive loss function \citep{chopra2005learning, hadsell2006dimensionality}, which minimises the Euclidean distance between embeddings from ``true" input pairs, and maximises it for ``false" input pairs. True and false pairs are automatically generated from the training data through a process known as self-supervision.

Formally, the network maps a window of ultrasound $w_u$, and a window of audio $w_m$ (represented as MFCC features), to two low dimensional embeddings $v_u$ and $v_m$:
\begin{equation}
    \begin{split}
        \psi(w_u; \theta) &\rightarrow v_u \\
        \phi(w_m; \eta) &\rightarrow v_m
    \end{split}
\end{equation}

Where $\psi$ and $\phi$ are non-linear transformations with parameters $\theta$ and $\eta$. The network then calculates the Euclidean distance $d$ between the embeddings:
\begin{equation}
    d = || v_u - v_m ||_2
\end{equation}

The learning objective is a contrastive loss $L$, which minimises the distance $d$ for true pairs (labelled $y=1$), and maximises it for false pairs (labelled $y=0$), for a number of training samples $N$: 
\begin{equation}\label{equation_E}
L(\theta, \eta) = \frac{1}{N} \sum^{N}_{n=1}
y_n d_n^2 + (1-y_n)\{max(1-d_n, 0)\}^2
\end{equation}

Once trained, the model can be used to calculate the Euclidean distance between a pair of ultrasound and audio windows. %Figure~\ref{fig:network} illustrates the model.

To find the synchronisation offset, we first need to specify the range of possible shifts (e.g., $\pm$1000 ms). Within this range, we use our model to identify the offset that minimises the mean Euclidean distance across shorter windows of the two signals. In practice, we discretise the range of possible shifts, rendering a discrete set of candidate offsets. Then, using Algorithm~\ref{alg:algorithm}, we calculate the mean euclidean distance for each of these candidates, and select the one with the smallest mean distance as our prediction. 

\begin{algorithm}[t]
\SetKwInput{KwInput}{Input} 
\SetAlgoLined
\KwInput{ultrasound, audio, and candidate offsets}
 \For{each candidate}{
    Apply candidate to utterance\\
    Create windows of ultrasound and audio \\
     \For{each window}{
        Calculate the distance between ultrasound and audio using UltraSync\\
    }
    Calculate the mean distance 
 }
 Select the offset with the smallest mean distance 
\caption{\label{alg:algorithm} Synchronisation algorithm}
\end{algorithm}
\begin{figure}[t]
\includegraphics[width=\columnwidth]{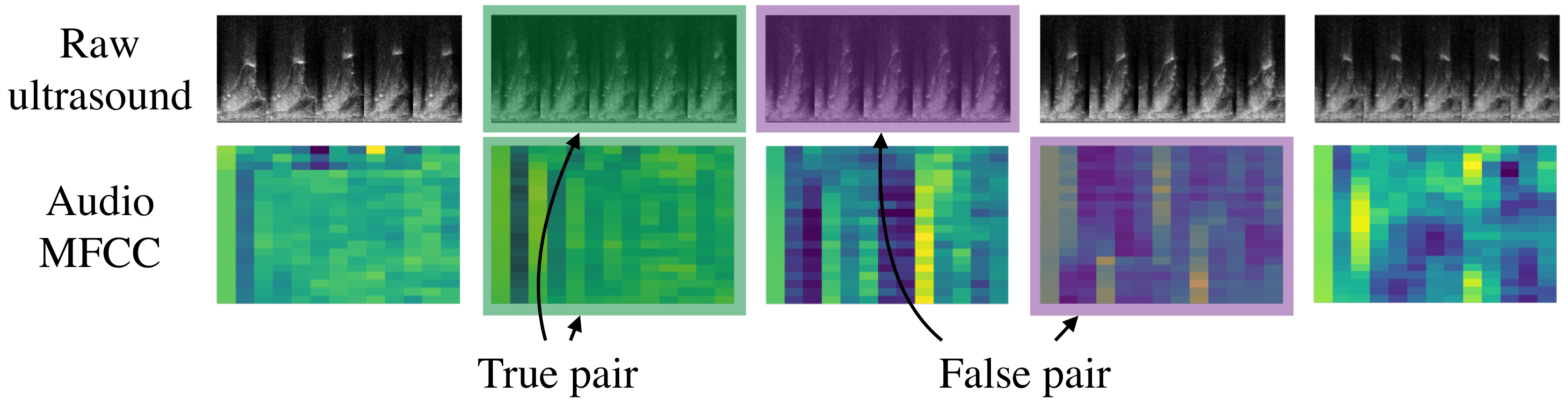}
\caption{We create training samples automatically using a self-supervision strategy. For each utterance, we create short windows of ultrasound and audio. True samples are corresponding pairs, and false samples are randomised pairings. Ultrasound frames are shown as raw reflection data.}
\label{fig:self-supervision}
\end{figure}

To train our model, all we require is a training set with correctly synchronised utterances, and from this dataset we automatically generate true and false pairs. From each utterance in the set, we generate multiple true pairs by creating short windows of ultrasound and corresponding audio and labelling them as true. To create false pairs, we simply randomise the pairings within each utterance, and label them as false. Figure~\ref{fig:self-supervision} illustrates the process of creating true and false samples.

\subsection{Experiment}

For this experiment, we used the UltraSuite and the TaL data. The datasets were recorded with speakers with different characteristics, in different environments, and using different equipment. We utilised such data to enable our model to accommodate different speakers groups, recording conditions, and ultrasound probe types. 
We split UltraSuite and TaL into training, validation, and testing subsets. We used the same data splits for UltraSuite as \citet{eshky2019synchronising}, and the same data splits for TaL as \citet{ribeiro2021tal} for comparability. We reiterate the data splits below.

% use the same sets from old paper or retrain without holding out sessions?
From UXTD, we used
speakers $[$7, 8, 12, 13, 26$]$ for validation, 
$[$30, 38, 43, 45, 47, 52, 53, 55$]$ for testing, and 
all remaining 45 speakers for training. 
From UXSSD, we used
speaker 1 and session `Mid' from speakers $[$2, 3, 4$]$ for validation,
speaker 7 and session `Mid' from speakers $[$5, 6, 8$]$ for testing, and 
all remaining speakers and sessions for training. 
From UPX, we used 
speaker 1 and session `BL3' from speakers $[$2-10$]$ validation,
speaker 15 and session `BL3' from speakers $[$11-14$]$ and $[$16-20$]$ testing, and 
all remaining speakers and sessions for training.
We used utterances containing words, non-words, sentences, isolated articulations, and conversations, and excluded utterance containing only coughs and swallowing motions.

From TaL1, we used
days $[$2, 3, 4$]$ for training, 
day 5 for validation, and 
days $[$1, 6$]$ for testing. 
From TaL80, we used 
speakers $[$1-49$]$ for training, 
speakers $[$50-65$]$ for validation, and
speakers $[$66-81$]$ for testing. 
We used read and spontaneous utterances and excluded swallow, silent, and whispered utterances.

A pre-processing step re-sampled audio at 22.05 KHz (using \emph{scipy interpolate}),
re-sampled ultrasound at 24fps (using \emph{samplerate resample}),
and resized ultrasound frames to $63 \times 138$ pixels (using \emph{skimage transform}).

We defined the sample window size as $\simeq$200ms long, 
calculated as $t = l/r$, where 
$t$ is the time window,
$l$ is the number of ultrasound frames per window (5 in our case), 
and $r$ is the ultrasound frame rate of the utterance (24 fps). 
For each utterance, we split the ultrasound into non-overlapping windows of 5 frames each. 
To create corresponding audio windows, we extracted MFCC features from the audio signal, with 13 cepstral coefficients, using a window length of $\simeq$20ms, calculated as $t/(l\times2)$, and a step size of $\simeq$10ms, calculated as $t/(l\times4)$. We chose MFCCs as they are one of the most frequently used representations in the speech processing literature, and have been shown to work for lip video synchronisation \citep{chung2016out}. 
We created true and false training samples using the process outlined in Figure~\ref{fig:self-supervision},
and generated as many false pairs as true ones for a balanced set.

The hyper-parameters of our network are shown in Table~\ref{table:parameters_table}. We pooled all training data and trained a single model using the Adam optimiser \citep{kingma2015adam}, with a learning rate of 0.001, a batch size of 64 samples, and for 20 epochs. We implemented learning rate scheduling, which reduced the learning rate by a factor of 0.1 when the validation loss plateaued for 2 epochs.

\begin{table}[t]
\centering
\resizebox{1\columnwidth}{!}{%
\begin{tabular}{@{}cccccc@{}}
\toprule
\textbf{Stream} & \textbf{Conv} & \textbf{Conv} & \textbf{Conv} & \textbf{FC} & \textbf{FC}\\
\midrule
Ultrasound & $23\times5\times5$ & $64\times5\times5$ & $128\times5\times5$ & $64$ & $64$\\
$5\times63\times138$ & $\times2$ pool & $\times2$ pool & $\times2$ pool &  & \\
\midrule
Audio & $23\times3\times3$ & $64\times3\times3$ & $128\times3\times3$ & $64$ & $64$\\
$1\times20\times30$ &  & $\times2$ pool & $\times2$ pool &  & \\
\bottomrule
\end{tabular}
}%
\caption{
Each stream had 3 convolutional (Conv) layers followed by 2 fully-connected (FC) layers.
FC layers had 64 units each.
For Conv layers, we specify the number of filters and their receptive 
field size as ``num $\times$ size $\times$ size" 
followed by the max-pooling down-sampling factor.  
Each layer was followed by batch-normalisation then ReLU activation. 
Max-pooling was applied after the activation function.}
\label{table:parameters_table}
\end{table}
%

%dim_visual	
%0	5
%1	63
%2	138
%dim_audio	
%0	1
%1	20
%2	13
%output_size	64
%optimiser	"Adam"
%num_epochs	20
%learning_rate	0.001

%experiment_id	"20200930-16hr02m39s"
%
%train	
%loss	0.18633430695905104
%accuracy	0.7166555122900949
%
%val	
%loss	0.18730739454726497
%accuracy	0.713250300120048
%
%test	
%loss	0.19551929994474387
%accuracy	0.6930193976518632

Upon convergence, the model achieved 
0.19 training loss, 
0.19 validation loss, and 
0.20 test loss, 
and by placing a threshold of 0.5 on predicted distances, the model achieved
71.7\% binary classification accuracy on training samples, 
71.3\% on validation samples, and
69.3\% on test samples.

\subsection{Evaluation and results}
\label{sec:model_results}
Next, we followed Algorithm~\ref{alg:algorithm} to predict the offsets for the test utterances, using the same 24 candidates for UltraSuite as \citet{eshky2019synchronising}, and using the same 25 for TaL as \citet{ribeiro2021tal}. 
%These are 24 candidates for UltraSuite and 25 for TaL.

To evaluate the predictions, we computed the discrepancy between the model prediction and the true offset as: 
\begin{equation}
disc = prediction - truth  
\end{equation}

Because hardware synchronisation was correct for UltraSuite and TaL, we treat it as $truth$. 
We consider the prediction to be correct if it falls between $lower$ and $upper$ thresholds: 
\begin{equation}
lower < disc < upper 
\end{equation}

Based on the new threshold defined in Section~\ref{sec:human_experiment}, we define two \textbf{accuracy scoring boundaries}:
\begin{enumerate}
    \item \textbf{Hard}: $lower=-$125ms and $upper=$45ms
    \item \textbf{Soft}: $lower=-$185ms and $upper=$90ms
\end{enumerate}
The hard scoring boundary is the same one used in previous work on lip synchronisation \citep{chung2016out} and ultrasound synchronisation \citep{eshky2019synchronising, ribeiro2021tal}. 
However, in Section~\ref{sec:human_experiment}, we found these thresholds to be too strict for ultrasound, and so we also present results using the soft scoring boundary. 

%subset	n	hard	soft	disc_mean	disc_std
%uxtd	455	64.6	74.5	123	392
%uxssd	396	88.9	95.7	12	146
%upx	651	93.7	98.3	0	90
%tal1	452	98.7	99.8	0	26
%tal80	3129	95.7	98.2	-8	54
%all	5083	92.4	96	7	140
%UltraSuite	1502	83.6	90.4	41	242
%Tal	3581	96.1	98.4	-7	51

\begin{table}[t]
\centering
\resizebox{0.83\columnwidth}{!}{%
\begin{tabular}{lrrrr@{}l}
\toprule
\multicolumn{1}{l}{\textbf{Subset}} & 
\multicolumn{1}{r}{\textbf{N}} & 
\multicolumn{1}{r}{\textbf{Hard}} & 
\multicolumn{1}{r}{\textbf{Soft}} & 
\multicolumn{2}{r}{\textbf{Discrepancy}}\\
\midrule
\multicolumn{6}{c}{\emph{UltraSuite: child data}} \\ 
\midrule
UXTD & 455 & 64.6\% & 74.5\% & 123 $\pm$ &~392 ms\\
UXSSD & 396 & 88.9\% & 95.7\% & 12 $\pm$ &~146 ms\\
UPX & 651 & 93.7\% & 98.3\% & 0 $\pm$ &~~~90 ms\\
& 1502 & 83.6\% & 90.4\% & 41 $\pm$ &~242 ms\\
\midrule
\multicolumn{6}{c}{\emph{TaL: adult data}} \\ 
\midrule
TaL1 & 452 & 98.7\% & 99.8\% & 0 $\pm$ &~~~26 ms\\
TaL80 & 3129 & 95.7\% & 98.2\% & -8 $\pm$ &~~~54 ms\\
 & 3581 & 96.1\% & 98.4\% & -7 $\pm$ &~~51 ms\\
\midrule
All & 5083 & 92.4\% & 96.0\% & 7 $\pm$ &~140\\
\bottomrule
\end{tabular}
}%
\caption{
Results by dataset. We show the accuracy using hard and soft scoring boundaries, and the mean and standard deviation of the discrepancy in milliseconds.
Performance on adult data (TaL) is better than on child data (UltraSuite).
}
\label{table:results-by-dataset}
\end{table}

%subset	n	hard	soft	disc_mean	disc_std
%A	914	92	97.7	3	107
%B	58	86.2	98.3	14	165
%C	186	94.6	97.3	11	105
%D	340	54.4	65.6	164	445
%F	4	100	100	-20	19
%aud	2979	95.8	98.2	-8	54
%xaud	432	97.7	99.3	-2	36
%spo	18	94.4	100	-9	32
%cal	152	98.7	100	-3	18
%all	5083	92.4	96	7	140

\begin{table}[t]
\centering
\resizebox{0.95\columnwidth}{!}{%
\begin{tabular}{lrrrr@{}l}
\toprule
\multicolumn{1}{l}{\textbf{Utterance Type}} & 
\multicolumn{1}{r}{\textbf{N}} & 
\multicolumn{1}{r}{\textbf{Hard}} &
\multicolumn{1}{r}{\textbf{Soft}} & 
\multicolumn{2}{r}{\textbf{Discrepancy}}\\
\midrule
\multicolumn{6}{c}{\emph{UltraSuite: child data}} \\ 
\midrule
Words & 914 & 92.0\% & 97.7\% & 3 $\pm$ &~107 ms\\
Non-words & 58 & 86.2\% & 98.3\% & 14 $\pm$ &~165 ms\\
Sentence & 186 & 94.6\% & 97.3\% & 11 $\pm$ &~105 ms\\
Articulatory & 340 & 54.4\% & 65.6\% & 164 $\pm$ &~445 ms\\
Conversation & 4 & 100\% & 100\% & -20 $\pm$ &~~~19 ms\\
%All & 1502 & 83.6\% & 90.4\% & 41 $\pm$ &~242 ms\\
\midrule
\multicolumn{6}{c}{\emph{TaL: adult data}} \\ 
\midrule
Read & 2979 & 95.8\% & 98.2\% & -8 $\pm$ &~~~54 ms\\
Read shared & 432 & 97.7\% & 99.3\% & -2 $\pm$ &~~~36 ms\\
Spontaneous & 18 & 94.4\% & 100\% & -9 $\pm$ &~~~32 ms\\
Calibration & 152 & 98.7\% & 100\% & -3 $\pm$ &~~~18 ms\\
%All & 3581 & 96.1\% & 98.4\% & -7 $\pm$ &~~51 ms\\
\bottomrule
\end{tabular}
}%
\caption{
Results by utterance type. 
We show the accuracy using hard and soft scoring boundaries, and the mean and standard deviation of the discrepancy in milliseconds.
Articulatory utterances contain isolated phones and are the most challenging. 
In contrast, performance is high on utterances containing natural variation in speech, 
such as words, sentences, conversations, read text, and spontaneous speech.  
}
\label{table:results-by-type}
\end{table}

Table~\ref{table:results-by-dataset} shows the results by dataset. The model correctly synchronises 92.4\% of utterances according to the hard scoring boundary and 96.0\% of utterances according to the soft scoring boundary.
The overall discrepancy is 7 $\pm$ 140 ms. 
Performance on TaL is better than on UltraSuite.
On child data (UltraSuite), the model achieves an overall hard accuracy of 83.6\%, a marginal improvement of 0.7\% over \citet{eshky2019synchronising}, and achieves a soft accuracy of 90.4\%. 
On adult data (TaL), the model achieves an overall hard accuracy of 96.1\%, a marginal reduction of 1.6\% over \citet{ribeiro2021tal}, and achieves a soft accuracy of 98.4\%. 
While these differences are small, they make intuitive sense.   
UltraSuite was recorded during speech therapy sessions in noisy environments, and the audio contains the speech of both therapists and patients. 
TaL on the other hand, was recorded in a hemi-anechoic chamber to eliminate background noise, and the audio and ultrasound always corresponded to the same speaker, resulting in much better overall quality.
Therefore, it is unsurprising that training on TaL improves the performance on UltraSuite, while training on UltraSuite reduces the performance on TaL.

Table~\ref{table:results-by-type} shows the results by utterance type. 
Performance according to both scoring boundaries is highest on utterances containing natural variation in speech, such as words, sentences, read text, and spontaneous speech. This result is consistent with the results from \citet{eshky2019synchronising}. 
Articulatory utterances, on the other hand, contain isolated phones (e.g.\ “sh”), and therefore lack natural variation in speech, which makes them more challenging to automatically synchronise. Nonetheless, the model correctly synchronises 54.4\% of these utterances according to the hard scoring boundary, and 65.6\% of the utterances according to the soft scoring boundary. 

Non-word stimuli are designed to elicit phones in different contexts from patients, but are not real English words (e.g.\ ``p apa epe opo"). To some extent, these utterances also lack natural variation in speech. According to the hard scoring boundary, 86.2\% of these utterances are correctly synchronised, which is lower than the accuracy achieved for words and sentences. However, using the slightly more flexible soft scoring boundary, 98.3\% of these utterances are considered correctly-synchronised, which slightly exceeds performance on words and sentences. At a first glace, this result seems surprising, but considering that many of these utterances contain repetitions of the same non-word, it is possible that the model is able to identify periodic landmarks in the utterances, and synchronise them to an adequate level, if not as precisely as it synchronises words and non-words. 

%\subsection{Summary and discussion}
To summarise, in this section we presented our approach for automatically synchronising ultrasound and audio. We introduced two scoring boundaries based on the detection thresholds from Section~\ref{sec:human_experiment}, and showed how to use them to evaluate our model. Results are consistent with previous work, demonstrating that performance is highest on utterances exhibiting natural variation in speech. 
TaL is of better quality than the UltraSuite data, and it is therefore unsurprising that the model achieves higher performance on TaL than on UltraSuite. 
Training a single model on the pooled TaL and UltraSuite data slightly reduces the performance on TaL and slightly increases it on UltraSuite, compared to previous research. 
In the next section, we evaluate our model's performance on the out-of-domain Cleft data. 
\section{Synchronising the Cleft data}
\label{sec:cleft_sync}

In this section, we test the performance of our system on the out-of-domain Cleft data. As described in Section~\ref{sec:data}, hardware synchronisation for the Cleft data was perceived as inadequate by the speech and language therapists who recorded it. Because correct synchronisation is not available for this data, we are unable to automatically evaluate model performance as we did in the Section~\ref{sec:model}. Instead, we utilise the judgement of experienced ultrasound users in a second perceptual experiment, which we describe below\footnote{This experiment was certified according to the Informatics Research Ethics Process (ref no 2019/43362).}. 

\subsection{Experiment}

\begin{table}[t]
\centering
\resizebox{1\columnwidth}{!}{%
\begin{tabular}{@{}cccc@{}}
\toprule
\textbf{Participant ID} & \textbf{Profession} & \textbf{Native English} & \textbf{Dialect}\\
\midrule
1 & SLT & Yes & Scottish \\
2 & SLT & Yes & Non-British \\
3 & Speech Scientist & Yes & British other \\
4 & Speech Scientist & Yes & Scottish \\
5 & SLT & Yes & Scottish \\
6 & SLT & Yes & Non-British \\
\bottomrule
\end{tabular}
}%
\caption{Details of the participants.}
\label{table:par_2}
\end{table}

The experimental setup is similar to that in Section~\ref{sec:human_experiment} 
with some differences which we outline below. 
We recruited a number of experienced ultrasound tongue imaging users,
giving them pairs of videos containing ultrasound tongue imaging and the corresponding audio, 
and asking them to choose the videos which they perceived to be better synchronised.
Each pair of videos were identical apart from the synchronisation offset. For one of the videos, we used the \textbf{original hardware synchronisation offset}. For the majority of utterances, this was perceived as inadequate by the speech and language therapists who collected the data. For the other video, we used the \textbf{offset predicted by our model}. The order of the videos was randomised and the source of the offset for each video was not shown to participants. 
This setting allowed us to measure the percentage of utterances for which the model improved synchronisation.

As in Section~\ref{sec:human_experiment}, the experiment was computer-based, and the videos were displayed on the participants' screens. The overall experiment lasted 30-40 minutes, and participants were allowed to complete it over multiple sessions.  
We gave the participants the option to play the videos at three speeds: 1.0$\times$, 0.5$\times$, and 0.25$\times$, and required them to play each video at least once and up to 6 times at any speed.
The participants could only move to the next pair of videos after submitting a judgement.

We asked the following question: ``In which of the two videos are the audio and tongue motion better synchronised, A or B?". Unlike the experiment in Section~\ref{sec:human_experiment}, we gave the participants only 2 choices: ``Video A", and ``Video B", and asked them to chose at random if they perceived no difference, or if the synchronisation in both videos was equally poor. In this setting, preference would approach 50\% if all choices were at random, and 100\% if one method was always preferred. To qualify, each participant was required to be a fluent English speaker, and either a speech and language therapist or a speech scientist with experience working with ultrasound tongue imaging. We recruited 6 participants whose details we outline in Table~\ref{table:par_2}.

\subsection{Data preparation}
The Cleft dataset contains 1441 samples of approximately 4.1 hours of audio in total. We evaluated only a subset of this data. We focused on evaluating spoken utterances (these are types A, B, and C in Table~\ref{table:cleft-summary}) and excluded ``swallows" (type E) which have almost no audible content. 
We evaluated utterances recorded during assessment, and excluded therapy utterances as they tend to be much longer and tend deviate from the prompt. Because the model was only trained on midsagittal utterances with the tip of the tongue to the right, we excluded utterances recorded in the coronal orientation.  
The duration of recordings in the dataset range from 2.4 to 40 seconds, with a mean of 10.3 seconds and a standard deviation of 5.1 seconds. We placed a threshold of $<=$15 seconds on utterances to evaluate, thereby excluding the tail of longer utterances.
We further excluded all utterances where the difference between the offset predicted by our model and the hardware offset fell within the undetectable range.

As we did in the first experiment, we listened to a small sample of recordings from each speaker to assess the audio quality. Out of the 29 speakers, we excluded 8 speakers who repeatedly deviated from the prompts and had the most interventions from therapists, because these kind utterances would distract our evaluators from the main task. 
We used the following speakers (9 female and 12 male): 3, 5, 7, 11, 14, 15, 16, 17, 18, 20, 21, 24, 25, 28, 30, 31, 32, 33, 34, 36, 39.

To apply our approach to the Cleft data, we needed to specify the range of offsets, as we did in Section \ref{sec:model}. 
We observed that for the majority of Cleft utterances, audio is advanced with respect to ultrasound, and so we considered a wider range of negative offsets than positive ones. The range we considered was [-1.75, 0.75] seconds with a step size of 45ms. This rendered 56 candidate offsets for the model to consider. We ran the model and reviewed a sample of predictions. We observed that the utterances with extreme offsets (largest and smallest) were poorly synchronised compared to utterances in the middle range. At this point, we had the option to either fine tune the range of candidate offsets, or sample utterance from the middle range. We chose to do the latter, randomly sampling 100 utterances of each utterance type (A, B, and C) within offsets [-1.5, 0.5], or a total of 300 utterances.

To test the reliability of participant choices, we added a small number of control utterances for which correct synchronisation was known. We used the UPX subset of UltraSuite, selecting 10 utterance with similar prompts to the Cleft dataset to obscure the origin of the utterances. We then created pairs of videos, which were identical apart from the synchronisation offset. For one of the videos, we use the correct hardware synchronisation offset and for the other, we added a detectable error of -305 ms for half of the utterances and 180 ms to the other half. All participants evaluated this same subset of 10 control utterances. 
In total, each participant evaluated 60 utterances, 50 Cleft samples and 10 control samples. 

\subsection{Results}
% accuracy, control: (91.7, (84.7, 98.7))
% preference, overall: (79.3, (74.8, 83.9))

\begin{figure}[t]
\includegraphics[width=\columnwidth]{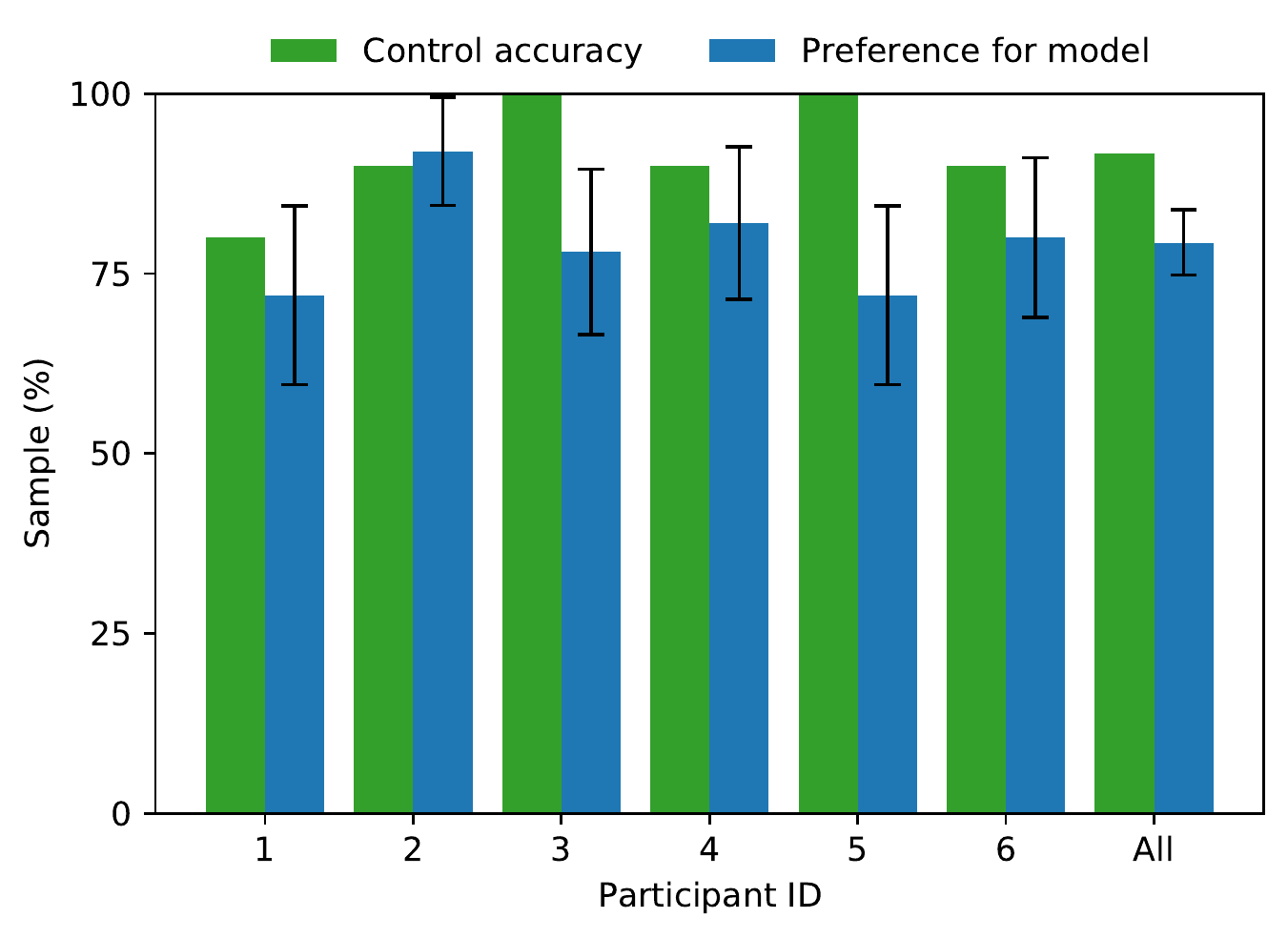}
\caption{The aggregate result and the result by participant. 
Green bars show high choice accuracy for control samples, and therefore, high participant reliability.
Blue bars show a strong preference for our model for Cleft samples. 
Confidence intervals are calculated for a binomial proportion.}
\label{fig:cleft}
\end{figure}

%A 100 78.0 (69.9, 86.1)
%B 100 86.0 (79.2, 92.8)
%C 100 74.0 (65.4, 82.6)

\begin{table}[t]
\centering
\resizebox{.85\columnwidth}{!}{%
\begin{tabular}{@{}ccccc@{}}
\toprule
\textbf{Utterance type} & 
\textbf{Type ID} & 
\textbf{N} & 
\textbf{Preference} &
\textbf{CI} \\
\midrule
Words       & A & 100 & 78.0\% & (69.9, 86.1) \\
Non-words   & B & 100 & 86.0\% & (79.2, 92.8) \\
Sentence    & C & 100 & 74.0\% & (65.4, 82.6) \\
\bottomrule
\end{tabular}
}%
\caption{Preference for our model, shown by utterance type. CI are 95\% binomial confidence intervals.}
\label{table:cleft_type}
\end{table}

\begin{table}[t]
\centering
\resizebox{.75\columnwidth}{!}{%
\begin{tabular}{@{}ccccc@{}}
\toprule
\textbf{Category} &
\textbf{P} & 
\textbf{N} & 
\textbf{Preference} &
\textbf{CI} \\
\midrule
SLT                 &  4    & 200 & 79.0\% & (73.4, 84.6) \\
Speech Scientist    &  2    & 100 & 80.0\% & (72.2, 87.8) \\
\midrule
Scottish        &  3    & 150  & 75.3\% & (68.4, 82.2) \\
British other   &  2    & 100  & 79.0\% & (71.0, 87.0) \\
Non-British     &  1    & 50   & 92.0\% & (84.5, 99.5) \\
\bottomrule
\end{tabular}
}%
\caption{Preference for our model, broken down by the participants' professions (top) and their dialects (bottom). P is the number of participants, while N is the number of samples. CI are 95\% binomial confidence intervals.}
\label{table:cleft_profession_lang}
\end{table}

Figure~\ref{fig:cleft} shows the aggregate result and the result by participant. 
Results show that participants are highly reliable, achieving an accuracy of 91.7\% with a confidence interval of (84.7, 98.7) for control utterances. 
%For control utterances, participants achieve an accuracy of 91.7\% with a confidence interval of (84.7, 98.7), which indicates that the participants are highly reliable. 
%
As for Cleft samples, participants preferred the model's prediction over hardware synchronisation 79.3\% of the time, with a confidence interval of (74.8, 83.9). 
We conduct a two-sided binomial test, achieving a p-value of $1.81 e^{-25}$ $<$ $0.001$, which indicates that the difference between the synchronisation methods is significant. We therefore have sufficient evidence that participants prefer the output from our model over the original hardware synchronisation. 

%We reject the null hypothesis that there is no difference between the two synchronisation methods. We have sufficient evidence that the output from our model is preferred. Participants prefer our model 79.2\% of the time.

Table~\ref{table:cleft_type} shows the preference for our model broken down by utterance type. According to participant choice, our model performs best on utterances of type ``non-words", followed by ``words" then ``sentences". As with the results in Section~\ref{sec:model_results}, this result may seem surprising at a first glance, as we expected performance to be higher on words and sentences because they exhibit slightly more natural variation in speech than non-words. However, the result is consistent with the Soft score calculated on in-domain data in Table~\ref{table:results-by-type}. Because many of the ``non-word" utterances contained repetitions of the same non-word (e.g., ``aka aka aka.."), it is possible that poor synchronisation was more obvious, and easier to detect by our participants.

Finally, we break the results down by the professions and dialects of the participants in Table~\ref{table:cleft_profession_lang}. Four of the participants were speech and language therapists (SLTs) and two were speech scientists. The results show no difference in model preference between the two groups. 
All participants were native English speakers: 3 Scottish, 2 non-Scottish British, and 1 non-British. 
The non-British speaker has a higher preference from our model, however due to the small sample size and the fact that the confidence intervals overlap with the non-Scottish British group, it is difficult to draw robust conclusions about the effects of dialect on model preference. 

To summarise, in this section we applied our model to the Cleft data and evaluated its performance with experienced ultrasound tongue imaging users. The participants showed a strong preference for our model's output over hardware synchronisation, which demonstrates our model's ability to generalise to data from a new domain.

%\section{Discussion}
%\label{sec:discussion}
% candidate set left up to the user. 

\section{Conclusion}
\label{sec:conclusion}
This paper addressed the problem of automatically synchronising ultrasound tongue imaging with speech audio. The two modalities are simultaneously-acquired; however, synchronisation information is not always correctly-captured at recording time, and is not always available for historical data. 

In Section~\ref{sec:human_experiment}, we presented a novel investigation of the synchronisation errors tolerance by expert ultrasound users, and found that thresholds for error detection are greater for ultrasound tongue imaging than for lip videos. We also found that sensitivity to synchronisation errors varies by participant, and that phones involving little tongue movement negatively correlate with a correct choice, while phones involving more tongue activity positively correlate with a correct answer. Findings from this experiment allowed us to define thresholds for detecting synchronisation errors in ultrasound.  

We then presented our approach for automatic synchronisation in Section~\ref{sec:model}, which utilises a self-supervised neural network to find the offset between ultrasound and audio in a given range. 
We defined two scoring boundaries for evaluating our model, a hard one and a soft one, based on the error thresholds we identified in our first perceptual experiment. We evaluated our approach in the first instance on in-domain data; a held-out subset of the data used to develop the model. Results are consistent with previous work, demonstrating that performance is highest on utterances exhibiting natural variation in speech. 
Our model achieved a higher performance on TaL than on UltraSuite, and
training a single model on the pooled TaL and UltraSuite data slightly reduced accuracy on TaL, while improving it on UltraSuite, compared to previous research. 

In Section~\ref{sec:data}, we introduced a novel resource, the Cleft dataset, which we collected with a new clinical subgroup, and for which hardware synchronisation proved unreliable. We applied our model to this data in Section~\ref{sec:cleft_sync} and evaluated it subjectively with expert users in a second perceptual experiment. We found that users preferred the output of our model 79.3\% of the time, and that this result is statistically significant. These results demonstrate the strength of our model and its ability to generalise to new domains. 

\section{Discussion and future work}
\label{sec:discussion}
There are several avenues for future research.
In Section~\ref{sec:human_experiment}, we investigated whether lip thresholds hold for ultrasound, and this served as a good starting point for identifying suitable thresholds to use for evaluating our system in Section~\ref{sec:model}. In the future, we can use the thresholds that we have identified as a guide to a new experiment which tests more fine-grained offsets to find more precise error detection thresholds. 

%In Sections~\ref{sec:human_experiment} 
Furthermore, in Section~\ref{sec:human_experiment}, we explored the notion of synchronisation error detection, but did not explicitly address ``acceptable" error, simply because it depends on the end task. As discussed in Section~\ref{sec:sync-background}, speech and language therapists use ultrasound differently to phoneticians, and so different tasks may require different levels of synchronisation precision. One future direction is to examine the effect of synchronisation error on the performance of experts in a downstream task, such as correctly identifying covert articulation errors. 
Within this context, we could also investigate whether different types of speech errors affect the ability of expert users to detect a synchronisation error, and whether there is a difference between typical and disordered speech.

In Section~\ref{sec:cleft_sync}, our perceptual experiment revealed that experienced ultrasound users prefer the output of our system to hardware synchronisation. This indicates that we were able to improve synchronisation overall but does not tell us how good the automatic synchronisation was. Because rating and subjective scoring can be unreliable, ascertaining whether the automatic synchronisation was done to an acceptable level is best conducted in the context of a downstream task, as proposed above. 

In Section~\ref{sec:model}, we trained our model on raw ultrasound data. 
However, other ultrasound systems used within the speech community produce DICOM sequences, or video recordings of ultrasound already in transformed format (AVI, MP4). Future work can explore transforming our data first and then training the model directly on the transformed images to make it applicable to these other formats. 

We can also extend our work to coronal ultrasound data. Because our model was trained on midsagittal utterances with the tip of the tongue to the right, we did not apply it to coronal Cleft utterances. In the future, we can explore collecting coronal images, validating their hardware synchronisation, and using them to adapt our model to this different orientation.

One limitation of our approach, which we identified while preparing the experiment in Section~\ref{sec:cleft_sync}, is the need to specify the range of candidate offsets as input, by examining some samples of poorly synchronised data. This domain knowledge can restrict our ability to integrate the model into a data pre-processing pipeline. In the future, we will explore ways to eliminate the need to specify the range of offsets as input.

\section{License and distribution}
\label{sec:license}
This manuscript bears a CC-BY-NC-ND license.
We distribute the Cleft dataset as part of the UltraSuite repository\footnote{\label{fn:ultrasuite}\url{https://ultrasuite.github.io/data/cleft/}} under the Attribution-NonCommercial 4.0 Generic license CC-BY-NC 4.0, and release the UltraSync model\footnote{\label{fn:ultrasuite-tools}\url{https://github.com/aeshky/ultrasync}} under the Apache License v.2. 

\section*{Acknowledgements}
We thank the speech and language therapists, and speech scientists who took part in our perceptual experiments. 
We thank the children who took part in the ``visualising speech" project, and their parents for allowing us to share the data with the research community. 
This work was funded by EPSRC Healthcare Partnerships Programme, grant number EP/P02338X/1 (Ultrax2020: \ultraxurl), and Action Medical Research, grant number GN2544 (Visualising Speech).

\bibliography{references}

\end{document}